\documentclass[prd,reprint,groupedaddress,nofootinbib]{revtex4-2}

\usepackage{graphicx}
\usepackage{dcolumn}
\usepackage{bm}
\usepackage{mathtools}
\usepackage{amssymb,amsmath}
\usepackage[protrusion=true,expansion=true]{microtype}
\usepackage[dvipsnames]{xcolor}

\usepackage[colorlinks=true,linkcolor=blue,citecolor=blue,urlcolor=blue]{hyperref}

\definecolor{webred}{rgb}{0.75,0,0}

\begin{document}

\title{Hidden-flavour four-quark states in the charm and bottom region}

\author{Joshua Hoffer$^{1,2}$}
\email[]{joshua.hoffer@theo.physik.uni-giessen.de}
\author{Gernot Eichmann$^3$}
\email[]{gernot.eichmann@uni-graz.at}
\author{Christian S. Fischer$^{1,2}$}
\email[]{christian.fischer@theo.physik.uni-giessen.de}

\affiliation{$^1$Institut für Theoretische Physik, Justus-Liebig-Universität Gießen, 35392 Gießen, Germany}
\affiliation{$^2$Helmholtz Forschungsakademie Hessen für FAIR (HFHF), GSI Helmholtzzentrum
für Schwerionenforschung, Campus Gießen, 35392 Gießen, Germany}
\affiliation{$^3$Institute of Physics, University of Graz, NAWI Graz, Universitätsplatz 5, 8010 Graz, Austria}

\date{\today}

\begin{abstract}
We discuss the spectrum and the internal composition of ground and excited four-quark states in the charm
and bottom energy region. To this end we extend previous calculations within the framework of the relativistic
four-body Faddeev-Yakubovsky equation to include quantum numbers with $J^{PC}=0^{++}, 0^{-+}, 1^{--}, 1^{+-}$ and $1^{++}$
and study their internal composition in terms of heavy-light meson pairs, hadroquarkonia and diquark-antidiquark
clusters. We observe similar patterns in the charm and bottom energy region with different compositions of the
four-quark states depending on $J^{PC}$ quantum numbers. Most notably, we find that all states with $C\cdot P=+1$
are dominated by heavy-light meson contributions, whereas for axialvector states with $J^{PC}=1^{+-}$ including
the $Z_c(3900)$ we find a much more complicated picture depending on the flavour content. We systematically compare
our results for the spectrum with existing experimental results and provide predictions for future analyses.
\end{abstract}

\maketitle

\section{Introduction}\label{intro}

Starting out with the unexpected detection of a narrow state in the $J/\psi\pi^+\pi^-$
invariant mass spectrum by the Belle collaboration in 2003 \cite{Choi2003}, i.e., the
$\chi_{c1}(3872)$, an ever increasing number of exotic states has been identified in the
charmonium and bottomonium mass region by Belle, BaBar, BSE III and the LHC experiments
in the last two decades. Many of these `$XYZ$ states' cannot be accommodated for in the
conventional quark model for $Q\bar{Q}$ mesons (with $Q=c,b$) and are therefore generally referred
to as exotic hadrons. Some of these states carry electric charge, which can only be explained
assuming four valence (anti-)quarks with either hidden or open heavy flavour configurations,
i.e. $Q\bar{Q}q\bar{q}$ or $QQ \bar{q}\bar{q}, \bar{Q}\bar{Q}{q}{q}$ (with $q=u,d,s$).
Thus, four-quark states are considered to be good candidates to study the properties
of these exotic hadrons, see, e.g.,
\cite{Esposito2017,Lebed2017,Chen2016,Ali2017,Guo2018,Olsen2018,Liu2019,Brambilla2020}
for recent reviews.

A highly debated and unsettled property of four-quark states is their internal structure.
In most effective field theory and model approaches, one can
generally distinguish between three different \textit{a priori} assumptions regarding
a possible internal clustering. The first possibility is motivated by the experimental
observation of final states with a specific charmonium/bottomonium state and light hadrons.
This is known as the \textit{hadroquarkonium} picture \cite{Voloshin2008} and features
a tight heavy quark and antiquark ($Q\bar{Q}$) core which is surrounded by the light
$q\bar{q}$ pair. The second prominent possibility is the clustering of the constituents
into \textit{diquark-antidiquark ($\mathit{dq-\overline{dq}}$)} pairs which interact via
colour forces, see, e.g., \cite{Esposito2017} for a review. Finally, there is the
\textit{meson-molecule} picture, which is especially relevant for states close to
open-flavour thresholds. In this picture the constituents arrange in pairs of
$D^{(\ast)}\bar{D}^{(\ast)}$ or $B^{(\ast)}\bar{B}^{(\ast)}$ mesons which
interact via short- and/or long-range forces~\cite{Guo2018}.

On a general note, these three possibilities of internal clustering are not mutually
exclusive, i.e., experimental states may be a superposition of all three different
clusters with the 'leading' component possibly different on a case-by-case basis.
To thoroughly study this behaviour, it is important to develop theoretical approaches
to QCD that can deal with all three possibilities. A prominent approach is lattice QCD,
where interesting progress has been made in recent years,
see \cite{Prelovsek2010,Berlin2015,Lee2014,Padmanath2015, Francis2017,Francis2019,Prelovsek2023,Bicudo2023,Bicudo2023a}
and references therein. Recently, the  framework of functional methods has been generalized
to systematically investigate four-quark states with any $J^{PC}$ and flavour combination,
see~\cite{Eichmann2020} and references therein for a review on results in the light and
charm mass region.

In this work we reanalyse and confirm the findings for the hidden-charm four-quark states
in \cite{Wallbott2020}, i.e., states with quantum numbers $I(J^{PC}) = 0(1^{++})$,
$1(1^{+-})$ and $0(0^{++})$. Furthermore, we investigate the experimentally very
interesting vector ($1^{--}$) channel in the hidden-charm region, and discuss our findings
for possible pseudoscalar ($0^{-+}$) four-quark states in that region, where there are
currently no experimental exotic candidates. We also present our novel results for the
$bq\bar{q}\bar{b}$ hidden-bottom mass spectrum with the aforementioned quantum numbers.
Last but not least, we discuss a new method to investigate the internal structure of
four-quark states by considering how much each of the three internal clusters
contributes to the normalization of a given state \cite{Eichmann2022,Torcato2023}.

The paper is organized as follows: In Section~\ref{bse} we briefly introduce the
four-body Bethe-Salpeter equation (BSE) and discuss the technical details, i.e.,
truncation of the two-body interaction, construction of the physical basis and
the internal structure. In Sec.~\ref{results} we present and discuss our results
for the hidden-charm and hidden-bottom four-quark states, before we conclude
in Sec.~\ref{conclusions}.

\section{Setup}\label{bse}

\subsection{Four-quark Bethe-Salpeter equation}\label{general_stuff}

In this work we focus on heavy-light and heavy-heavy four-quark states with hidden flavour. We denote
their quark content by $Qq\bar{q}\bar{Q}$ with $q\in\{u,d,s,c,b\}$ and $Q\in\{c,b\}$.
Any bound state or resonance in QCD that has overlap with $Qq\bar{q}\bar{Q}$ appears as a pole in the
quark eight-point correlation function or scattering matrix $T^{(4)}$, which
satisfies the  scattering equation
\begin{equation}\label{eq: T-matrix}
T^{(4)} = K^{(4)} + K^{(4)}G_0^{(4)} T^{(4)}\,.
\end{equation}
Here, $K^{(4)}$ is the four-body interaction kernel and $G_0^{(4)}$ denotes a product
of four dressed (anti-)quark propagators; see \cite{Eichmann2016} for details.
From the pole residue of Eq.~(\ref{eq: T-matrix}) one obtains the homogeneous
four-quark BSE, written in compact notation as
\begin{equation}\label{eq: four-quark-bse}
\Gamma^{(4)} = K^{(4)}G_0^{(4)}\Gamma^{(4)}\,.
\end{equation}
Each multiplication implies an integration
over all loop momenta, and $\Gamma^{(4)}$ is the four-quark Bethe-Salpeter amplitude of a given state.

Eq.~\eqref{eq: four-quark-bse} is an eigenvalue equation for $K^{(4)}G_0^{(4)}$
with eigenvalues $\lambda_i(P^2)$, which depend on the total hadron momentum squared
$P^2\in\mathbb{C}$. If the condition $\lambda_i(P_i^2) = 1$ is satisfied, this
corresponds to a pole in the scattering matrix at $P_i^2=-M_i^2$. The index $i$
indicates whether the eigenvalue satisfying the condition corresponds to the ground
state ($i=0$), the first radial excited state ($i=1$), etc. Below a given meson-meson
threshold, $M_i$ is real and we have found a bound state. For a resonance, on the other hand,
this condition is only satisfied in the complex plane on a higher Riemann sheet. In principle the
homogeneous BSE is able to detect both bound states and resonances, where
contour deformations are required to calculate $\lambda_i(P^2)$ above the
lowest threshold \cite{Williams2019,Eichmann2019,Santowsky2022}.

\begin{figure}[t]
\includegraphics[scale=0.38]{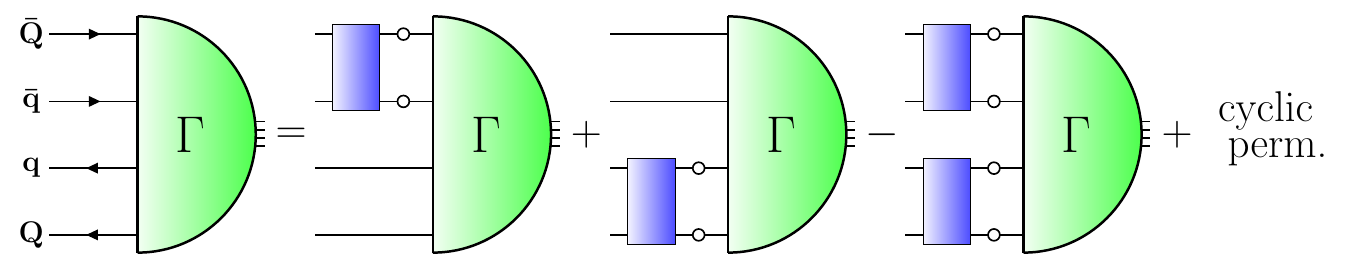}
\caption{\label{fig: bse}Four-quark BSE for a generic hidden-flavour $Qq\bar{q}\bar{Q}$ system in the $(12)(34)$ topology; the permutations $(13)(24)$ and $(14)(23)$ are not shown here. The green half-circles denote the Bethe-Salpeter amplitudes, blue boxes represent the two-body interaction kernels and the blobs denote fully-dressed quark propagators.}
\end{figure}

 The scattering kernel $K^{(4)}$ consists of irreducible \mbox{two-,}
 three- and four-body correlations.
In this work we primarily want to study the internal two-body clusters,
hence we neglect the three- and four-body forces.
The resulting kernel is then the sum of the two-body interactions
\begin{equation}\label{eq: two-body-kernel}
\tilde{K}^{(2)}G_0^{(4)} = \sum_{aa'}\left(K_a + K_{a'} - K_aK_{a'}\right)  ,
\end{equation}
where $a$ and $a'$ denote interactions between quark-(anti-) quark pairs and $aa'$
is one of three combinations $(12)(34)$, $(13)(24)$ and $(14)(23)$. These
correspond to the two-body interaction topologies for quark content $Qq\bar{q}\bar{Q}$:
Diquark-antidiquark $(Qq)(\bar{q}\bar{Q})$, meson-meson $(Q\bar{q})(q\bar{Q})$ and
hadro-quarkonium $(Q\bar{Q})(q\bar{q})$. Note that the last term on the r.h.s.
of Eq.~(\ref{eq: two-body-kernel}) is necessary to avoid overcounting \cite{Huang1975,Kvinikhidze1992,Heupel2012,Kvinikhidze2023}.
The resulting four-quark BSE with the kernel in Eq.~(\ref{eq: two-body-kernel}) is shown in Fig.~\ref{fig: bse}.

For the two-body kernels, we employ the rainbow-ladder truncation in combination
with the effective Maris-Tandy (MT) interaction~\cite{Maris1997,Maris1999}.
This setup models the combined effect of the dressed gluon propagator and
quark-gluon vertex and has been extensively applied to  meson,
baryon and four-quark phenomenology, see~\cite{Eichmann2016,Eichmann2020} for
recent reviews. The explicit form of the  interaction
can be found in Eq.~(3.96) of~\cite{Eichmann2016}; we use the scale parameter
$\Lambda = 0.72$~GeV tuned to  reproduce the pion decay constant
and fix the  shape parameter to $\eta^{\mathrm{MT}} = 1.8$.
The Dyson-Schwinger equation (DSE) for the quark propagator,
which is needed as an input for the BSE (cf.~Fig.~\ref{fig: bse}),
is solved using the same interaction.
The MT interaction is known to describe the  phenomenology of light mesons
in the pseudoscalar and vector channel reasonably well, whereas in the scalar
and axialvector channels it does not provide satisfactory results.
Its qualitative reliability can be judged from the meson masses
in Table~\ref{tab: meson-masses} obtained via the quark-antiquark BSE.

We work in the isospin symmetric limit,
i.e., $m_{\pi^\pm}=m_{\pi^0}$, $m_{D^\pm}=m_{D^0}$, $m_{B^\pm}=m_{B^0}$.
Throughout this paper we  use the abbreviation $n=u/d$ when referring to the light up and down quarks.
The $u/d$-quark mass is fixed by $m_\pi$ to $m_n=3.7$ MeV at a renormalization point $\mu=19$ GeV in a MOM scheme.
The strange and charm quark masses are chosen as $m_s=85$ MeV and $m_c=795$ MeV (in the same renormalisation scheme)
such that the sums $m_{D_s}+m_{D_s^\ast}$ and $m_{D}+m_{D^\ast}$ have $< 0.5\%$ deviation from the
sums of the respective experimental values \cite{Workman2022}.
The bottom quark mass is fixed to $m_b=3750$ MeV (again same scheme) such that the pseudoscalar and
vector $b\bar{b}$ masses and the sum $m_{B}+m_{B^\ast}$ matches experiment within $0.8\%$ relative error.
With these constraints set, the deviations between the theoretical and
experimental meson masses are then below $7\%$ in most channels.

\begin{table*}  \renewcommand{\arraystretch}{1.2}
\setlength{\tabcolsep}{0.5em} 
\centering
\begin{tabular}{c|ccc|ccc|ccc|ccc||cc}
& \multicolumn{3}{c|}{$0^{-+}$} & \multicolumn{3}{c|}{$1^{--}$} & \multicolumn{3}{c|}{$0^{++}$} & \multicolumn{3}{c||}{$1^{++}$} & $0^+$ & $1^+$\\[4pt]
 & PDG & $m_{\rm{RL}}$ & $\Delta m^{\rm{rel.}}$ & PDG & $m_{\rm{RL}}$ & $\Delta m^{\rm{rel.}}$ & PDG & $m_{\rm{RL}}$ & $\Delta m^{\rm{rel.}}$ & PDG & $m_{\rm{RL}}$ & $\Delta m^{\rm{rel.}}$ & $m_{0^+}$ & $m_{1^+}$ \\[1mm]
\hline\hline\rule{-0.9mm}{4mm}
$n\bar{n}$ &  $\pi/\eta\, ^\dagger$ &  $137$ & $0.0\%$ & $\rho/\omega$ & $736$ & $5.2\%$ & $f_0(1370)$ & $1370$ & $^\ddagger$ & $a_1$ & $898$ & $27.0\%$ & $809$ & $1006$ \\
$s\bar{n}$ &  $K$ & $501$ &$1.1\%$ & $K^\ast$ & $913$ & $0.1\%$ & $K_0^\ast(1430)$ & $1425$ & $^\ddagger$ & $K_1$ & $1110$ & $11.4\%$ & $1072$ & $1259$  \\
$s\bar{s}$ &  $-$ & $698$ &$-$ & $\phi$ & $1070$ & $5.0\%$ & $f_0(1500)$ & $1522$ & $^\ddagger$ &  $-$ & $1248$ & $-$ & $1266$ & $1412$ \\
$c\bar{n}$ &  $D$ & $1860$ &$0.4\%$ & $D^\ast$ & $2011$ & $0.1\%$ & $D_0^\ast$ & $2012$ & $14.1\%$ & $D_1$ &  $2174$ & $10.1\%$ & $2421$ & $2439$ \\
$c\bar{s}$ &  $D_s$ & $1937$ &$1.6\%$ & $D_s^\ast$ & $2124$ & $0.5\%$ & $D_{s0}^\ast$ & $2181$ & $5.9\%$ & $D_{s1}$ & $2273$ & $7.6\%$ & $2523$ & $2543$  \\
$c\bar{c}$ &  $\eta_c$ & $2803$ &$6.1\%$ & $J/\psi$ & $2992$ & $3.4\%$ & $\chi_{c0}$ & $3142$ & $8.0\%$ & $\chi_{c1}$ & $3154$ & $8.3\%$ & $3415$ & $3433$  \\
$b\bar{n}$ &  $B$ & $5310$ &$0.6\%$ & $B^\ast$ & $5375$ & $0.9\%$ & $-$ & $5550$ & $-$ & $B_1$ & $5802$ & $1.3\%$ & $6396$ & $6403$  \\
$b\bar{s}$ &  $B_s$ & $5425$ &$1.1\%$ & $B_s^\ast$ & $5487$ & $1.3\%$ & $-$ & $5680$ & $-$ & $B_{s1}$ &  $5909$ & $1.4\%$ & $6473$ & $6492$ \\
$b\bar{c}$ &  $B_c$ & $6232$ &$0.7\%$ & $-$ & $6302$ & $-$ & $-$ & $6538$ & $-$ & $-$ & $6655$ & $-$ & $7139$ & $7269$   \\
$b\bar{b}$ &  $\eta_b$ & $9421$ & $0.2\%$ & $\Upsilon$ & $9500$ & $0.4\%$ & $\chi_{b0}$ & $9759$ & $1.0\%$ & $\chi_{b1}$ & $9768$ & $1.1\%$ & $9915$ & $10394$  \\
\end{tabular}
\caption{\label{tab: meson-masses}
$Q\bar{q}$ mesons with quantum numbers $J^{PC} = \{0^{-+},1^{--},0^{++},1^{++}\}$ grouped according to their quark model classification.
We show the experimental candidates \cite{Workman2022}, the masses $m_{\rm{RL}}$ obtained
in our rainbow-ladder calculation, and the relative error of these to the masses given in the PDG (if the experimental state has been identified).
In the last two columns we also show the obtained rainbow-ladder masses for the corresponding $Qq$ diquarks with quantum numbers $J^P = \{0^+,1^+\}$.
All values are given in MeV.
$\dagger:$ The $\pi$ and the $\eta$ are mass degenerate in this work, since we neglect the strange component in the $\eta$ and the indirect
effect of the topological mass via octet-singlet mixing.
 $\ddagger:$ We do not consider the lightest scalar meson nonet as potential internal components, since they themselves are of four-quark
 nature \cite{Pelaez:2015qba,Eichmann2016a}. Instead, we resort to the scalar nonet with masses above 1 GeV.
}
\end{table*}

\subsection{Physically motivated four-quark amplitude}\label{subsec: physical basis}

In general, the four-quark Bethe-Salpeter amplitude is a direct product of Dirac (D),
colour (C) and flavour (F) parts. For  a given $J^{PC}$ it can be written as
\begin{equation}\label{eq: bse_w_dcf}
\Gamma^{(\mu\nu\ldots)}(p_1,p_2,p_3,p_4) = \Gamma_\mathrm{D}^{(\mu\nu\ldots)}(p_1,p_2,p_3,p_4)\otimes \Gamma_{\mathrm{C}}\otimes \Gamma_{\mathrm{F}}\ ,
\end{equation}
with $p_1,\ldots,p_4$ denoting the quark momenta. The $\mu\nu\ldots$ occur as Lorentz indices for states with higher spin $J$.
The colour and flavour part of the amplitude are straightforward to work out and discussed in
detail in the supplemental material of \cite{Wallbott2020}.

The Dirac part can be written as
\begin{equation}\label{eq: bse_w_d}
\Gamma_{\alpha\beta\gamma\delta}^{(\mu)}(p_1,p_2,p_3,p_4) = \sum_{i=1}^N f_i(\Omega)\,\tau_i^{(\mu)}(p_1,p_2,p_3,p_4)_{\alpha\beta\gamma\delta}\, ,
\end{equation}
where the $f_i(\Omega)$ are Lorentz-invariant dressing functions depending on
ten Lorenz invariant momentum variables $\Omega$ (see~\cite{Wallbott2019}
for details). The $\tau_i^{(\mu)}$ are the corresponding Dirac structures (with Dirac
indices $\alpha,\, \beta,\, \gamma\,, \delta$) and $N$ is the number of Dirac basis
elements. The full Dirac bases for $J=0$ ($N=256$) and $J=1$ ($N=768$) states are
collected in \citep{Eichmann2016a} and in the Appendix of \cite{Wallbott2019}.

Following the arguments in \cite{Eichmann2016a,Wallbott2020}, we note that
the amplitude dynamically develops two-body clusters in the three different
topologies mentioned in Section~\ref{general_stuff}. For heavy-light systems,
this is a heavy-light meson-meson  component ($\mathcal{M}_1$), a hadro-quarkonium component
($\mathcal{M}_2$) and a $\mathit{dq-\overline{dq}}$ cluster  ($\mathcal{D}$).
These clusters were found to heavily influence the four-body system, thus the
guiding idea is to express $\Gamma^{(\mu)}(p_1,p_2,p_3,p_4)$ in terms of these
internal two-body clusters.

\begin{figure}
\includegraphics[scale=0.4]{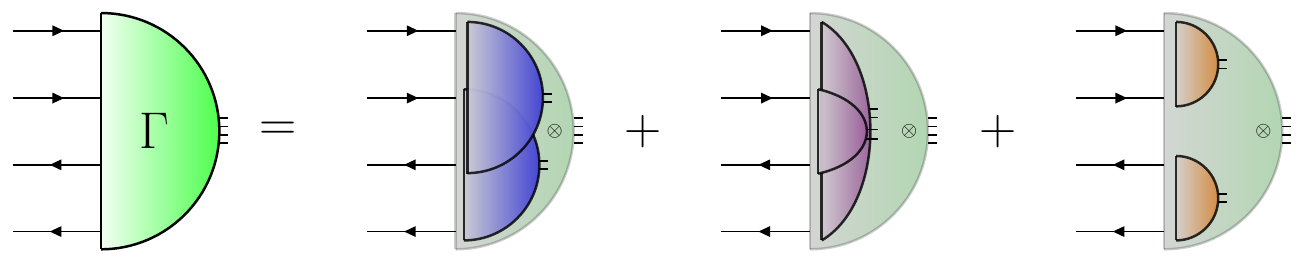}
\caption{\label{fig: physical_bse}Graphical representation of the Bethe-Salpeter amplitude in the physical basis.
The diagrams on the r.h.s. represent the direct product of the internal clusters
spanning the physical basis. The first diagram shows the  meson-meson
configuration ($\mathcal{M}_1$), with the heavy-light meson clusters depicted as blue half-circles. The second diagram
is the hadro-quarkonium contribution ($\mathcal{M}_2$), with individual components shown as
violet half-circles, and the last diagram is the diquark-antidiquark configuration
($\mathcal{D}$), where each diquark  is depicted by an orange half-circle. }
\end{figure}

Applying this idea to Eq.~(\ref{eq: bse_w_dcf}),
we construct a physically motivated basis for the Bethe-Salpeter amplitude by projecting onto
a subset of the full basis corresponding to the
dominant two-body clusters. For given quantum numbers $J^{PC}$, we draw on
existing information on the decay channels of experimental four-quark
candidates to identify the possible two-body clusters.
This in turn  fixes the Dirac tensors that enter in the amplitude.
A collection of the chosen
internal configurations used in this work is given in Table~\ref{tab: components}.
With all of the above, the amplitude in Eq.~\eqref{eq: bse_w_dcf} reduces to
\begin{equation}\label{eq: bse_physical}
\Gamma^{(\mu)}(\ldots) \approx\hspace*{-1.8em} \sum_{i\in\{\mathcal{M}_1,\mathcal{M}_2,\mathcal{D}\}}\hspace*{-1.8em} f_i(\Omega)\,\tau_i^{(\mu)}(\ldots)\otimes \tau_i^\mathrm{C}\otimes \tau_i^\mathrm{F}\, ,
\end{equation}
where the sum includes the dominant physical components for the different
interaction topologies in Table~\ref{tab: components}.
The resulting structure is visualized in Fig.~\ref{fig: physical_bse}.
We emphasize that this representation is different from the two-body framework used in \cite{Santowsky2022a},
as the sub-amplitudes depicted in Fig.~\ref{fig: physical_bse} are not actual two-body amplitudes
but rather specific Dirac-colour-flavour basis elements of the four-body amplitude 
that reflect its  internal clustering. 

In Eq.~\eqref{eq: bse_physical}, $\tau_i^\mathrm{C}$ is the attractive colour singlet structure corresponding to
each interaction topology, with $\mathbf{1}\otimes\mathbf{1}$ tensors for each of the two meson-meson topologies
and a $\mathbf{\bar{3}}\otimes \mathbf{3}$ tensor for the diquark-antidiquark topology (see Supplemental Material of \cite{Wallbott2020} for details). We relegate the inclusion of
repulsive colour-singlet channels to future work. The index $i$ on the flavour part $\tau_i^\mathrm{F}$ is
only used to construct the Dirac and colour part of the wavefunction
but is otherwise irrelevant for the calculation, as the interaction kernel described in Section \ref{general_stuff} is flavour-blind.

\begin{table}
\setlength{\tabcolsep}{0.5em} 
\centering
\begin{tabular}{ccc}
 & $I(J^{PC})$ & Physical components\\[3pt]
 &             & $f_0, \quad f_1, \quad f_2$    \\[4pt]
\hline
\rule{0pt}{1.\normalbaselineskip}
$cn\bar{n}\bar{c}$ & $0(0^{-+})$ & $D^*\bar{D}_1,\, \chi_{c0}\eta,\, \eta_{c} \tilde{f}_0$\\[4pt]
                   & $0(1^{--})$ & $D\bar{D}_1,\, \chi_{c0}\omega,\, J/\psi \tilde{f}_0$\\[4pt]
                   & $0(0^{++})$ & $D\bar{D},\, J/\psi\omega,\, S_{c}S_{c}$\\[4pt]
                   & $0(1^{++})$ & $D\bar{D}^\ast,\, J/\psi\omega,\, S_{c}A_{c}$\\[4pt]
                   & $1(1^{+-})$ & $D\bar{D}^\ast,\, J/\psi\pi,\, S_{c}A_{c}$\\[4pt]
  \hline
  \rule{0pt}{1.\normalbaselineskip}
$bn\bar{n}\bar{b}$ & $0(0^{-+})$ & $B^*\bar{B}_1,\, \chi_{b0}\eta,\, \eta_{b} \tilde{f}_0$\\[4pt]
                   & $0(1^{--})$ & $B\bar{B}_1,\, \chi_{b0}\omega,\, \Upsilon \tilde{f}_0$\\[4pt]
                   & $0(0^{++})$ & $B\bar{B},\, \Upsilon\omega,\, S_{b}S_{b}$\\[4pt]
                   & $0(1^{++})$ & $B\bar{B}^\ast,\, \Upsilon\omega,\, S_{b}A_{b}$\\[4pt]
                   & $1(1^{+-})$ & $B\bar{B}^\ast,\, \Upsilon\pi,\, S_{b}A_{b}$\\[4pt]

\end{tabular}
\caption{\label{tab: components} Physical content of the Bethe-Salpeter amplitude
for $cn\bar{n}c$ and $bn\bar{n}\bar{b}$ configurations, where $n$ stands for light $u/d$ quarks.
Scalar and axialvector diquarks are denoted by $S_{c/b}$ and $A_{c/b}$,
respectively, where the subscript characterizes the heavy quark that is paired with
the light quark. The $\tilde{f}_0$ here denotes the $f_0(1370)$.
The notation $f_0,f_1,f_2$ for the physical components corresponds to Eq.~(\ref{eq: dirac_bse_w_poles})
and is used again to display results in subsection \ref{subsec: internal}.
}
\end{table}

\subsection{Two-body poles}\label{subsec: physical poles}

Let us now take a closer look at the Dirac part of the amplitude.
One can express the quark momenta $p_1,\ldots,p_4$ by three relative momenta
$k,q,p$ and the total hadron momentum $P$ using the  relations
\begin{align}\label{eq: parameter_rewriting}
\begin{matrix}
p_1 = \frac{k+q-p}{2}+\sigma_1 P\, , \;\; & p_3 = \frac{-k+q+p}{2} + \sigma_3 P\, ,\\[6pt]
p_2 = \frac{k-q+p}{2}+\sigma_2 P\, , \;\; & p_4 = -\frac{k+q+p}{2} + \sigma_4 P\,.
\end{matrix}
\end{align}
Here,  $\sigma_1,\ldots,\sigma_4$ are (quark-)momentum partitioning parameters
which can be used to maximize the calculable domain for the bound-state mass
by optimally distributing the total hadron momentum among the  quarks. In the rest frame of the four-quark state, $P^\mu = (0,0,0,iM)$ is imaginary whereas the relative momenta $p,\, q,\, k$ are real. Therefore, the $p_i^2$ obtained from Eq.~(\ref{eq: parameter_rewriting}) describe different parabolas in the complex plane which are limited by the nearest quark singularities, which translates into an upper limit for $M$. An optimized choice of momentum partitionings then maximizes the mass range $M$ for which the BSE can be solved.

Using the relations in~\cite{Eichmann2015}, one can group the Lorentz-invariant momentum
variables $\Omega = \{q^2,p^2,k^2,\ldots\}$ in multiplets of the permutation group
$S_4$. This yields a singlet variable $S_0 = (k^2+q^2+p^2)/4$
carrying the momentum scale, a doublet  containing the internal two-body
clusters, and two triplets, thus totalling to $3+3+2+1 = 9$ momentum variables,
plus $P^2 = -M^2$. Following~\cite{Eichmann2016a,Wallbott2019},
the leading momentum dependence of the dressing functions $f_i$ beyond
the singlet variable $S_0$ comes from the two-body clusters,
whose poles are dynamically generated when solving the equation.
We therefore pull  out these poles  from the $f_i$ and,
to reduce computational effort,
assume that the remainder only depends on $S_0$.
The resulting Dirac part of the amplitude then reads
\begin{equation}\label{eq: dirac_bse_w_poles}
	\Gamma^{(\mu)}_{\alpha\beta\gamma\delta} (k,q,p,P) = \sum_{i=0}^2 f_i(S_0) P_{ab}^i P_{cd}^i \, \tau_i^{(\mu)}(k,q,p,P)_{\alpha\beta\gamma\delta} \, ,
\end{equation}
where the two-body poles of the amplitude are given by
\begin{equation}\label{eq: poles}
P_{ab}^i P_{cd}^i = \frac{1}{\big(p_{ab}^+\big)^2 + m_{ab}^2}\,  \frac{1}{\big(p_{cd}^+\big)^2 + m_{cd}^2}\, .
\end{equation}
Here, $p^+_{ab} = p_a + p_b$ is the momentum
of the  meson or diquark with mass $m_{ab}$ in a given topology $(ab)(cd) =(13)(24)$, $(14)(23)$ or $(12)(34)$.
The sum in Eq.~(\ref{eq: dirac_bse_w_poles}) runs over the physical components given in Tab.~\ref{tab: components}.
These occur in the diagrammatic topologies discussed in Fig.~\ref{fig: physical_bse}.
For example, for a four-quark state with $I(J^{PC})=0(1^{++})$
the $\mathcal{M}_1$ topology has clusters $m_{13} = m_D$ and $m_{24}=m_{\bar{D}^\ast}$,
the $\mathcal{M}_2$ cluster  $m_{14} = m_{J/\psi}$ and $m_{23}=m_{\omega}$, and
the $\mathcal{D}$ topology  $m_{12} = m_{S_{cq}}$ and  $m_{34}=m_{A_{\bar{q}\bar{c}}}$. 
On the other hand, the first two states of Tab.~\ref{tab: components} would have a single contribution ($f_0$)
	in $\mathcal{M}_1$ topology, two contributions ($f_{1,2}$) in $\mathcal{M}_2$ and none in $\mathcal{D}$.
In general, the respective meson and diquark masses
are calculated from their rainbow-ladder BSEs
as described above and compiled in Table~\ref{tab: meson-masses}.

\begin{figure*}[t]
	\includegraphics[scale=0.54]{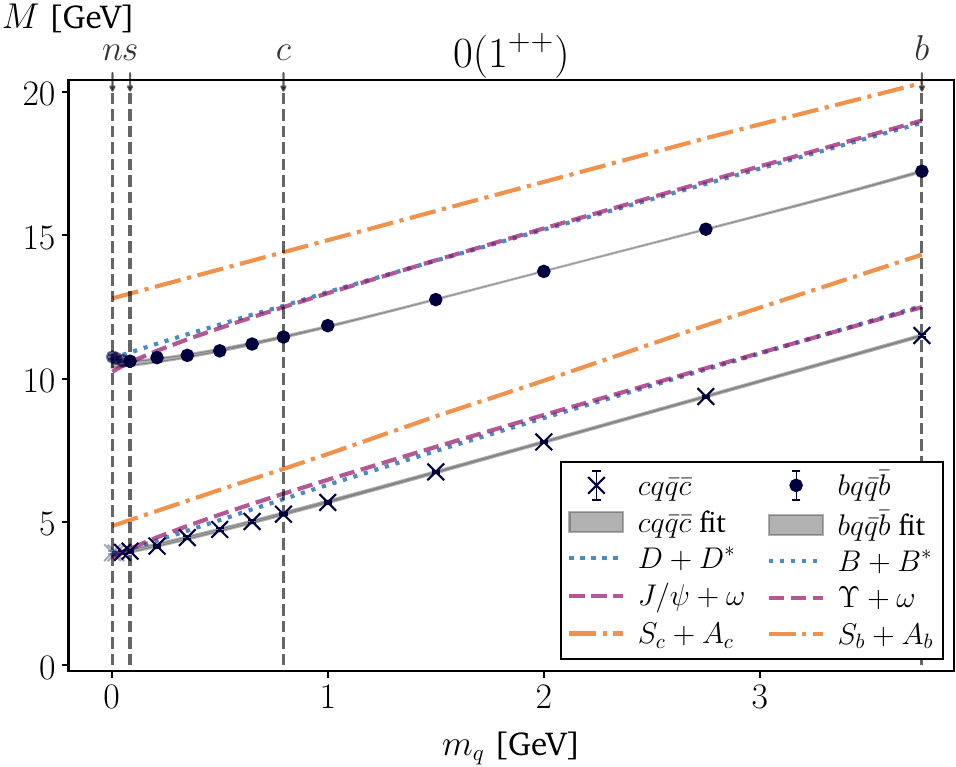}\hfill
	\includegraphics[scale=0.54]{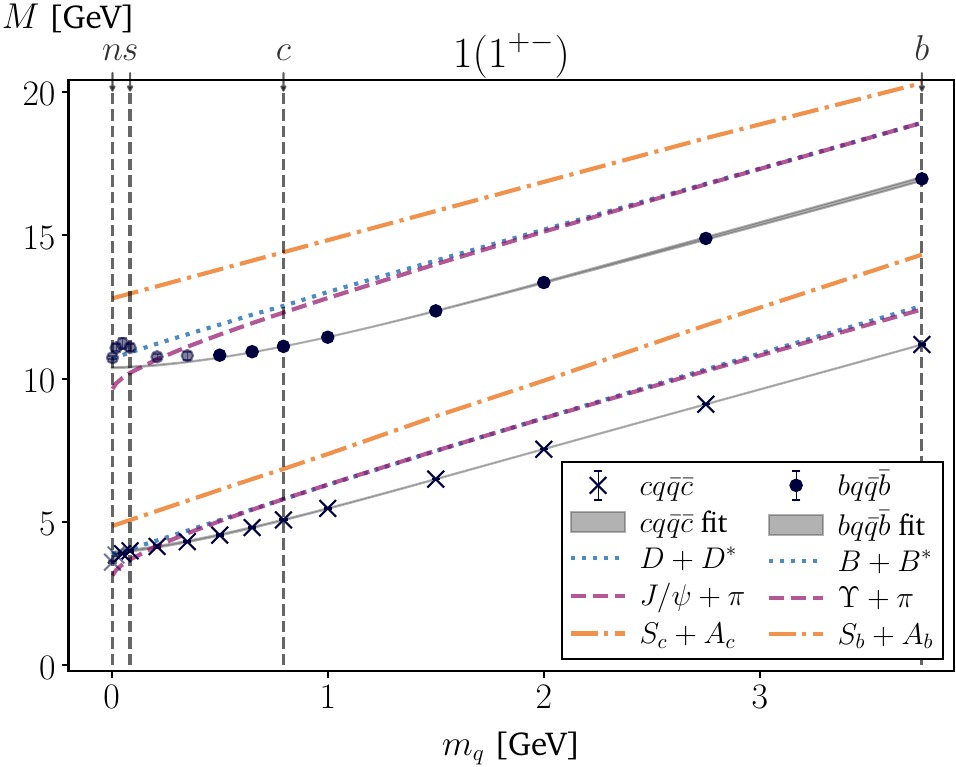}
	\caption{\label{fig: old} Current-mass evolution of the $cq\bar{q}\bar{c}$ (crosses) and $bq\bar{q}\bar{b}$ ground states (dots)
		in the $0(1^{++})$ and $1(1^{+-})$ channels. The grey vertical dashed lines mark the position of
		the $q=n,s,c,b$ current-quark masses, where $n=u/d$. The dotted, dashed and dash-dotted curves represent
		the meson-meson and diquark-antidiquark thresholds for the charm and  bottom four-quark system, respectively.
		The colours of the thresholds comply with Fig.~\ref{fig: physical_bse}, i.e., blue for the $\mathcal{M}_1$, purple for the
		$\mathcal{M}_2$ and orange for the $\mathcal{D}$ cluster. The grey bands
		are the fits to the data.}
\end{figure*}

Eq.~(\ref{eq: poles}) introduces two-body poles in
the integration domain of the r.h.s. of Eq.~(\ref{eq: four-quark-bse}),
thus restricting the range of $P^2=-M^2$ where the BSE can be solved directly.
This can be somewhat remedied by optimizing the quark momentum partitioning
parameters  in Eq.~(\ref{eq: parameter_rewriting}) which split the hadron momentum $P$ amongst the quarks.
On the other hand, the momenta in the denominator of Eq.~(\ref{eq: poles}), which are linear combinations of Eq.~(\ref{eq: parameter_rewriting}), also form complex parabolas which are limited by the meson and diquark poles. To this end, it is advantageous to relate the $\sigma_i$
to new parameters $\eta$, $\zeta$ and $\chi$, where $\eta$ corresponds to the meson-meson
topology ($\mathcal{M}_1$), $\zeta$ to the hadro-quarkonium ($\mathcal{M}_2$) and $\chi$ to the
$\mathit{dq-\overline{dq}}$ ($\mathcal{D}$) cluster:
\begin{equation}\label{eq: relation_parameters}
\begin{matrix}
\sigma_1 = \frac{1}{2}(\eta+\zeta+\chi -1)\, , & \sigma_2 = \frac{1}{2}(-\eta-\zeta+\chi + 1)\, ,\\[4pt]
\sigma_3 = \frac{1}{2}(\eta-\zeta-\chi + 1)\, , & \sigma_4 = \frac{1}{2}(-\eta+\zeta-\chi + 1)\, .
\end{matrix}
\end{equation}
The choice
\begin{align*}\label{eq: eta_zeta_chi}
\eta = \frac{m_{13}}{m_{13}+m_{24}}, \;\;
\zeta = \frac{m_{14}}{m_{14}+m_{23}}, \;\;
\chi = \frac{m_{12}}{m_{12}+m_{34}}
\end{align*}
then maximizes the value of $M$ to be
the lowest sum of masses of the individual
physical components in Table~\ref{tab: components}, e.g., $m_{D}+m_{D^\ast}$ for the $\chi_{c1}(3872)$.
Here we also remedy a slight inconsistency in the previous works~\cite{Wallbott2019,Wallbott2020},
where the quark momentum partitioning parameters were chosen as $\sigma_i = \frac{1}{4}$ in the pole terms~(\ref{eq: poles})
but set to their optimal values in the rest of the equation. As a result,
the maximum value of $M$ did not exhaust its full range,
which resulted in the need for extrapolating the eigenvalue curves over a large momentum range
causing a large extrapolation error. In the present work we overcome this limitation, thereby reducing the
extrapolation error considerably.

To calculate the eigenvalues above the thresholds, in principle one needs to employ contour deformation techniques
or elaborate analytic continuations~\cite{Eichmann2019,Santowsky2022}. As the former would present
an enormous technical challenge in the four-body approach, we relegate it to future work
and analytically continue the  eigenvalues  on the real axis using the Schlessinger-point method~\cite{Schlessinger1968}, see
Sec.~\ref{subsec: error} for details. Thus, if masses above thresholds are quoted
they merely serve as a rough estimate for the real part of the corresponding resonance locations.
In many cases, however, the resulting ground-state masses are below all thresholds and thus no analytic continuation is needed.

\section{Results}\label{results}

In the following we present  results for hidden-charm and hidden-bottom four-quark states
with quantum numbers $0(0^{++})$, $0(1^{++})$, $1(1^{+-})$, $0(1^{--})$, $0(0^{-+})$. The
first three of these were already investigated in the hidden-charm sector in \cite{Wallbott2019,Wallbott2020}.
Here we extend the calculations to the bottom region and also present novel results for the vector and pseudoscalar channels.
We first present our mass evolution curves in section \ref{subsec: mass-evol} and then discuss our results
for the physical spectrum in the charm and bottom energy region in sections \ref{subsec: charm} and \ref{subsec: bottom}.
The internal composition of our states is then the topic of section \ref{subsec: internal}.

\subsection{Mass evolution curves}\label{subsec: mass-evol}

We first discuss the charm- and bottom-like ground states in the $I(J^{PC})=0(1^{++})$, $1(1^{+-})$
and $0(0^{++})$ channels. While  the $0(1^{++})$ channel has established experimental four-quark candidates
only  in the charm region~\cite{Workman2022}, e.g.,
$\chi_{c1}(3872)$ and $\chi_{c1}(4140)$, the $1(1^{+-})$ channel features four-quark
candidates both in the charm region (e.g., the $Z_c(3900)$) and the bottom region, namely the $Z_b(10610)$ and $Z_b(10650)$.
The situation in the $0(0^{++})$ channel is not yet settled in the literature, but the
existing exotic candidates also only occur in the charm region.

\begin{figure}[!b]
\includegraphics[scale=0.54]{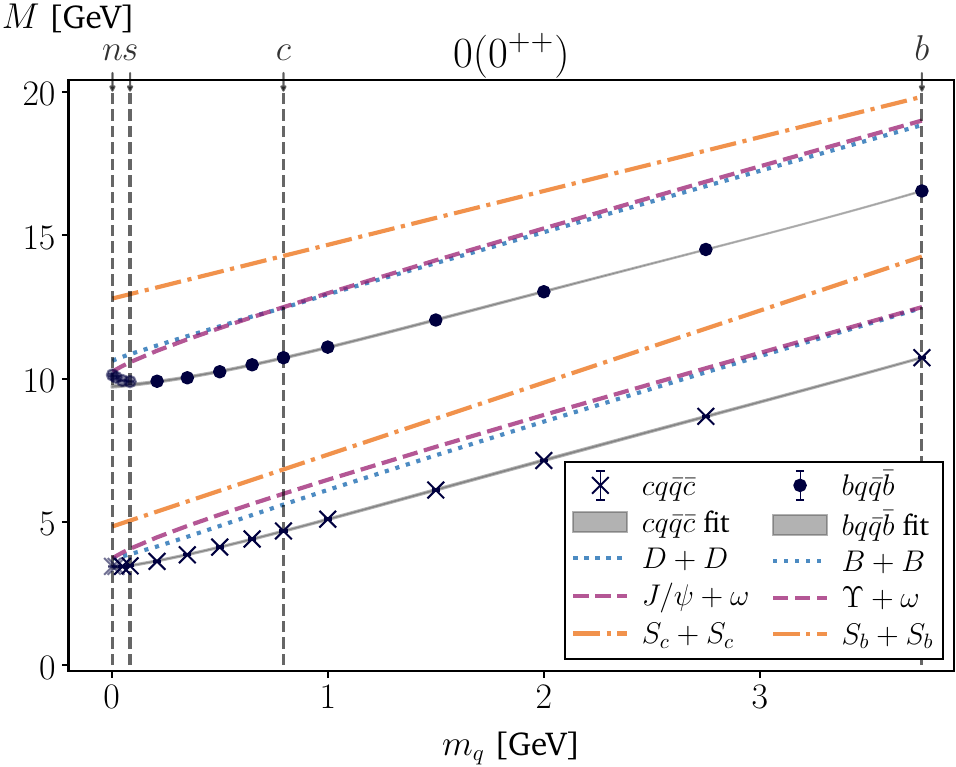}
\caption{\label{fig: old-1} Current-quark mass evolution of the $cq\bar{q}\bar{c}$ and $bq\bar{q}\bar{b}$ ground states
in the $0(0^{++})$ channel; see Fig.~\ref{fig: old} for details.}
\end{figure}

\begin{figure*}[t]
\includegraphics[scale=0.54]{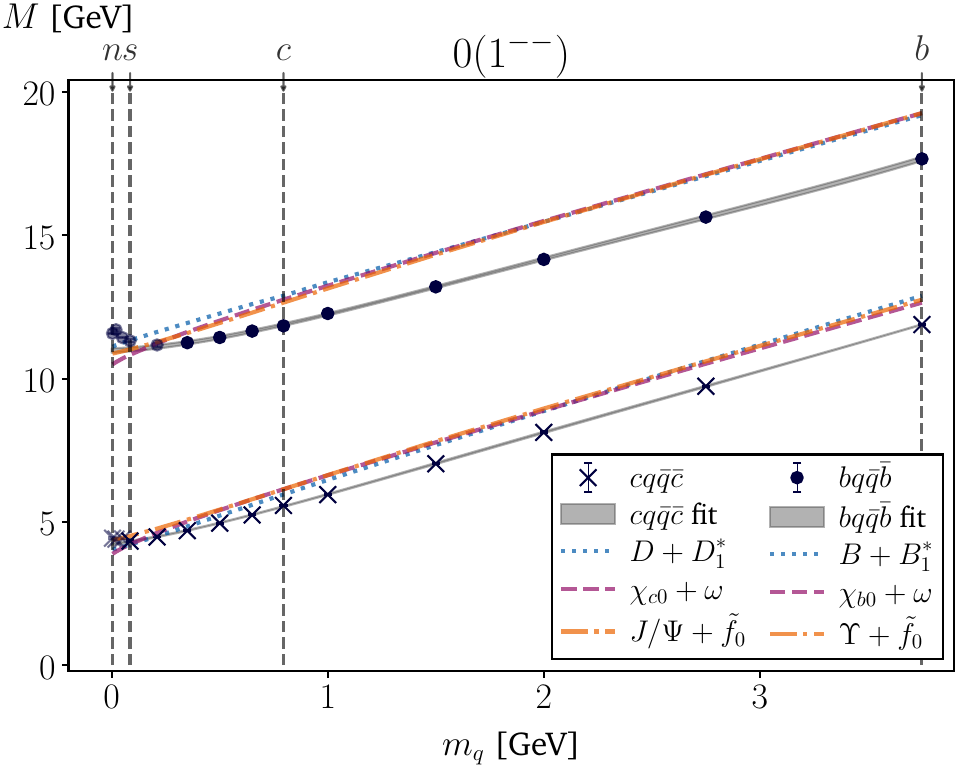}\hfill
\includegraphics[scale=0.54]{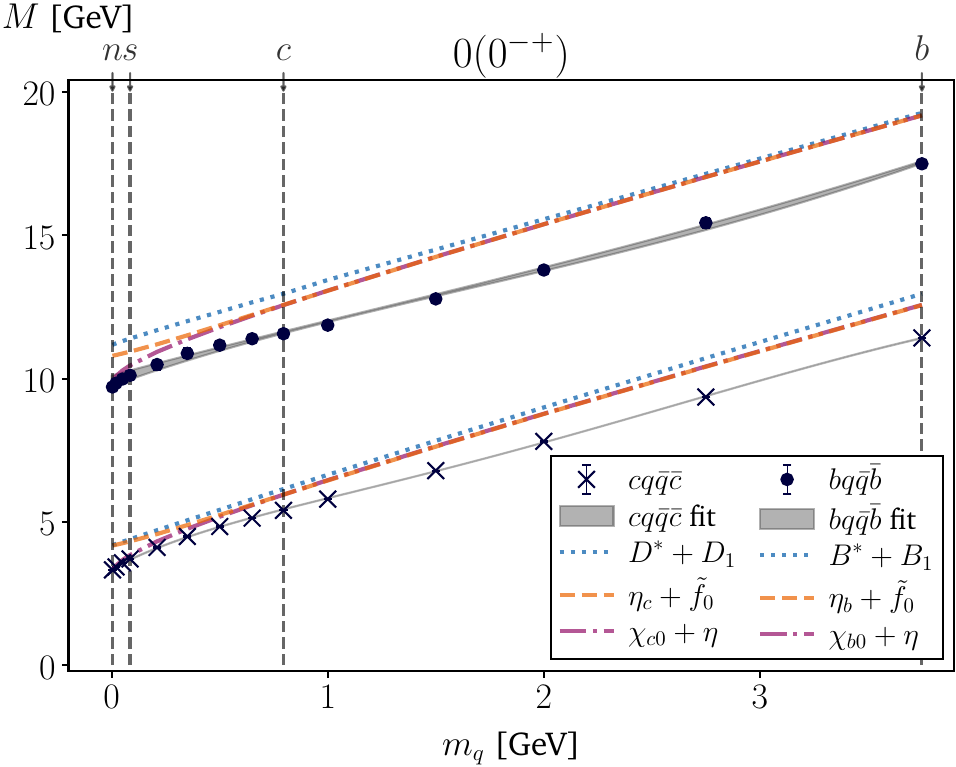}
\caption{\label{fig: new} Current-quark mass evolution of the $cq\bar{q}\bar{c}$ and $bq\bar{q}\bar{b}$ ground states
in the  $0(1^{--})$ and $0(0^{-+})$ channels; see Fig.~\ref{fig: old} for details.}
\end{figure*}

We display the results of our calculation for these three channels in Figs.~\ref{fig: old} and~\ref{fig: old-1},
where we show the mass evolution curve (MEC) of the four-quark state with fixed heavy-quark pair
$Q\bar{Q}=c\bar{c}$ (lower group of curves) and $b\bar{b}$ (upper group of curves). The quark pair $q\bar{q}$ is varied from the
bottom mass $m_b$ (rightmost vertical dashed line) to the light quark mass $m_n$ (leftmost
vertical dashed line). The results for  charm-like states are marked
by  crosses and those for  bottom-like states by dots.
The dotted, dashed and dash-dotted curves show the quark-mass evolution of the
two-body thresholds (cf.~Table~\ref{tab: components}). We show the MECs for the first radial excited states in Appendix~\ref{subsec: excited_states}.

We find that for increasing current-quark masses
the  four-quark states become more deeply bound with respect to the lightest meson-meson threshold.
For quark masses $m_q \geq m_c$, the MECs become approximately linear both in
the charm and  bottom region. However, below the charm mass we observe an upwards bending of the MEC
when it approaches the lowest meson-meson threshold in the system, which
is stronger for the $bq\bar{q}\bar{b}$ compared to the $cq\bar{q}\bar{c}$ states.
We note  that such a bending is also observed in the MECs for the two-body heavy-light
states in our framework. In the pseudoscalar channel, the MECs bend downwards for $m_q\simeq m_n$, resembling the observed behaviour of the MECs for the two-body $q\bar{q}$ states in the pseudoscalar (and scalar) channel, see, e.g., right figure in Fig.~3.10 in \cite{Eichmann2016}. Since the masses from the aforementioned two-body MECs serve as input for the four-quark state calculations, the similarity in the behaviour of the MECs could be interpreted as a first indication that the states in the pseudoscalar channel have a strong hadro-quarkonium component. A more detailed discussion about the internal structure can be found in Sec.~\ref{subsec: internal}.

In Tables~\ref{tab: masses} and~\ref{tab: excited masses}
we quote the masses of all  ground and excited states  calculated in this work.
To this end, we employ fits of the form
\begin{equation}\label{eq: fit}
M(m_q) = \sqrt{a+b \,m_q+c \,m_q^2}
\end{equation}
to the MECs at quark masses reasonably far away from the two-body thresholds.
These fits are shown in Figs.~\ref{fig: old} and~\ref{fig: old-1} by the grey bands.
The data not taken into account in the fits are depicted slightly opaque.
The error of the masses, given in the brackets in Table~\ref{tab: masses}, is then determined
by combining the width of the bands (i.e., the error of the fit function $M(m_q)$),
with  the extrapolation error as described in Appendix~\ref{subsec: error}.
To facilitate comparisons with the literature,
we also list the resulting binding energies $E_B = M - M_{\mathrm{th}}$ with respect to
our \textit{calculated} lightest heavy-light meson-meson threshold $M_\mathrm{th}$ in both tables. For the pseudoscalar channel, we instead use the dominant hadro-quarkonium threshold, see Sec.~\ref{subsec: internal}.
Note that these thresholds do not necessarily coincide with the experimental ones
given that our meson masses are  calculated in the same framework as our four-quark masses.

We now move on to the novel results of this work, i.e., the vector $0(1^{--})$ and pseudoscalar $0(0^{-+})$
four-quark states. For the $0(1^{--})$ channel there are many established exotic candidates in the
charm and bottom mass region~\cite{Workman2022}: the $\psi(4230)$, $\psi(4360)$, $\psi(4660)$,
$\Upsilon(10753)$, $\Upsilon(10860)$, and $\Upsilon(11020)$. By contrast, as of today there are
no experimental exotic candidates in the pseudoscalar channel. Thus, we can only compare our
results with the experimental situation in the vector channel and cross-check our obtained results with theoretical predictions for the pseudoscalar channel.

In contrast to the other three quantum numbers, our ``physical basis'' for the vector and pseudoscalar
four-quark states does not contain a diquark topology but rather a second $\mathcal{M}_2$ cluster (cf.
Table~\ref{tab: components}). The reason for this is that, on one hand, the diquark clusters always constitute
the highest two-body thresholds in our system and are generally found to be almost negligible for hidden-flavour
four-quark states. On the other hand, the construction of $S$-wave $\mathit{dq-\overline{dq}}$ pairs
in the $0(1^{--})$ and  $0(0^{-+})$ channels would require including pseudoscalar and vector diquarks,
which are strongly suppressed compared to their (``good'') scalar and (``bad'') axialvector diquark counterparts
(see Appendix~\ref{app: new amplitudes} for details).

Our results for the ground states of the vector and pseudoscalar states are displayed in Fig.~\ref{fig: new}.
In both channels, the behaviour of the MECs are very similar to the ones described before, i.e., the results are
again affected by the meson-meson thresholds for light quark masses.

As mentioned in Sec.~\ref{general_stuff}, we calculate both the four-quark ground states and radial excitations
from Eq.~(\ref{eq: four-quark-bse}). The resulting masses of the radial excitations are given in Table~\ref{tab: excited masses}.
Note that because the lowest two-body thresholds are identical for the ground and excited states, the eigenvalue curves need to be
extrapolated much further in some cases and thus the errors for these states increase.
Most of the excited states are therefore also unbound resonances.

As noted before, the binding energies in Tables~\ref{tab: masses} and~\ref{tab: excited masses} are determined
with respect to our calculated lightest heavy-light meson-meson thresholds. In particular, in the vector and
pseudoscalar channels the thresholds depend on the calculated masses of the scalar and axialvector $Q\bar{q}$ states,
which may differ from experiment by a couple of percent (cf. Sec.~\ref{general_stuff} and Tab.~\ref{tab: masses} for details).

\subsection{Charm spectrum}\label{subsec: charm}

\begin{figure*}[t]
\includegraphics[width=0.5\textwidth]{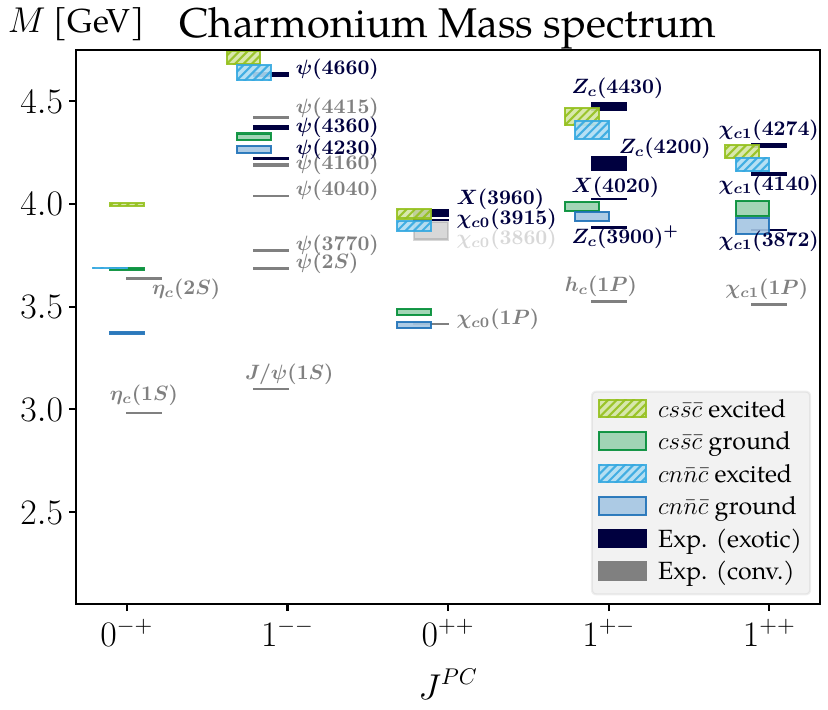}\hfill \includegraphics[width=0.5\textwidth]{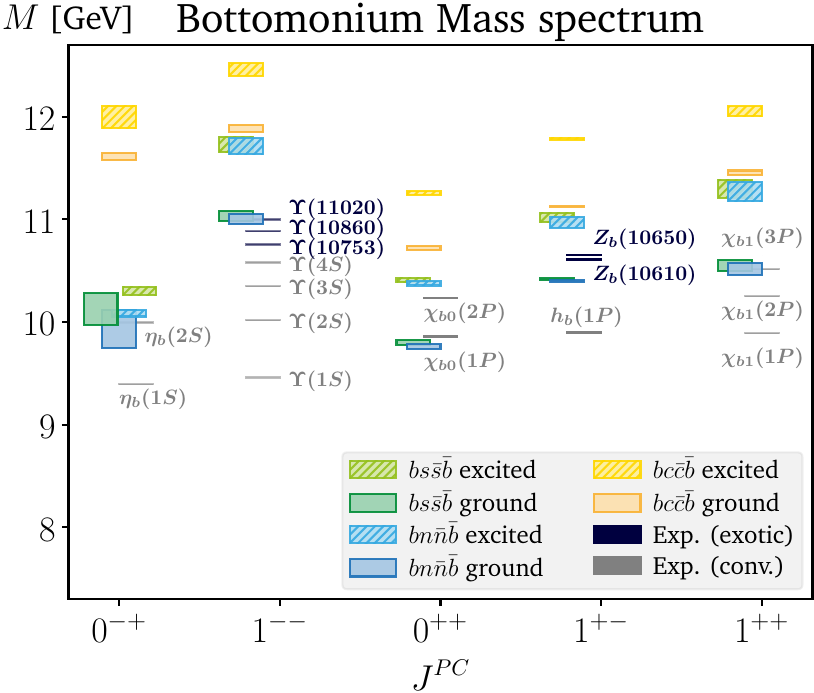}
\caption{\label{fig: spectrum}
Hidden-charm (\textit{left}) and hidden-bottom spectrum (\textit{right}) for the ground and first excited four-quark states compared to experiment~\cite{Workman2022}.
The coloured boxes are our results, where the height of the boxes stands for the error of the extracted masses.
The gray and black boxes are the PDG masses (real parts of the pole positions) for conventional and exotic hadrons, respectively.
The pale gray coloured states are not yet well established.}
\end{figure*}

In the left panel of Fig.~\ref{fig: spectrum} we compare our results for the masses in Tables~\ref{tab: masses} and~\ref{tab: excited masses}
to the experimental spectrum in the charmonium region. We find that the
$cn\bar{n}\bar{c}$ and $cs\bar{s}\bar{c}$ states, depicted by blue and green boxes, respectively, lie quite close together
in most channels. This closeness can be attributed to the plateau-like
behaviour of the MECs in Figs.~\ref{fig: old}--\ref{fig: new} for small quark masses. The only exception is the pseudoscalar channel,
where the MEC bends downwards at small quark masses.

\begin{table*}
\setlength{\tabcolsep}{0.60em} 
\centering
\begin{tabular}{c|c|c|c|c|c|c|c|c|c|c}
 & \multicolumn{2}{c|}{$0(0^{++})$} & \multicolumn{2}{c|}{$0(1^{++})$} & \multicolumn{2}{c|}{$1(1^{+-})$} & \multicolumn{2}{c|}{$0(1^{--})$} & \multicolumn{2}{c}{$0(0^{-+})$} \\[4pt]
\rule{0pt}{1.\normalbaselineskip}& $M$ & $E_B$ & $M$ & $E_B$ & $M$ & $E_B$ & $M$ & $E_B$ & $M$ & $E_B$ \\
\hline\hline\rule{-0.9mm}{4mm}
\rule{0pt}{1.\normalbaselineskip}$cn\bar{n}\bar{c}$ & $3.41(2)$ & $-0.31(2)$ & $3.89(4)$ & $0.02(4)$ & $3.94(2)$ & $\textcolor{gray}{0.07(2)}$ & $4.27(2)$ & $\textcolor{gray}{0.23(2)}$ & $3.37(1)$ & $\textcolor{gray}{0.09(1)}$ \\[4pt]
$cs\bar{s}\bar{c}$ & $3.47(1)$ & $-0.40(1)$ & $3.98(4)$ & $-0.08(4)$ & $3.99(2)$ & $-0.07(2)$ & $4.33(2)$ & $\textcolor{gray}{0.12(2)}$ & $3.68(1)$ & $-0.16(1)$ \\[4pt]
$bn\bar{n}\bar{b}$ & $9.77(2)$ & $-0.85(2)$ & $10.52(6)$ & $-0.17(6)$ & $10.40(1)$ & $-0.28(1)$ & $11.01(5)$ & $-0.11(5)$ & $9.9(2)$ & $-0.9(2)$ \\[4pt]
$bs\bar{s}\bar{b}$ & $9.80(2)$ & $-1.05(2)$ & $10.55(6)$ & $-0.36(6)$ & $10.42(1)$ & $-0.49(1)$ & $11.03(5)$ & $-0.30(5)$ & $10.1(2)$ & $-0.8(2)$ \\[4pt]
$bc\bar{c}\bar{b}$ & $10.72(2)$ & $-1.74(2)$ & $11.46(2)$ & $-1.08(2)$ & $11.13(0)$ & $-1.41(0)$ & $11.89(3)$ & $-1.00(3)$ & $11.61(4)$ & $-0.96(4)$ \\[4pt]
\end{tabular}
\caption{\label{tab: masses} Ground-state masses for the hidden-charm ($cq\bar{q}\bar{c}$) and hidden-bottom ($bq\bar{q}\bar{b}$) states in GeV.
For completeness we also display the binding energies $E_B$ with respect to the  lightest (calculated) heavy-light meson-meson threshold in each channel except for the $0^{-+}$, where we take the $\chi_{c0}\eta$ in the charm and the $\eta_b \tilde{f}_0$ in the bottom region, which are the most relevant thresholds in the system; the ``binding energies''
for resonant particles above the threshold are shown in grey. The error given in the brackets is the combination of the extrapolation error
and the error of the fit from Eq.~(\ref{eq: fit}).}
\vspace{1em}
\setlength{\tabcolsep}{0.60em} 
\centering
\begin{tabular}{c|c|c|c|c|c|c|c|c|c|c}
 & \multicolumn{2}{c|}{$0(0^{++})$} & \multicolumn{2}{c|}{$0(1^{++})$} & \multicolumn{2}{c|}{$1(1^{+-})$} & \multicolumn{2}{c|}{$0(1^{--})$} & \multicolumn{2}{c}{$0(0^{-+})$} \\[4pt]
\rule{0pt}{1.\normalbaselineskip}& $M$ & $E_B$ & $M$ & $E_B$ & $M$ & $E_B$ & $M$ & $E_B$ & $M$ & $E_B$ \\
\hline\hline\rule{-0.9mm}{4mm}
\rule{0pt}{1.\normalbaselineskip}$cn\bar{n}\bar{c}$ & $3.89(2)$ & $\textcolor{gray}{0.17(2)}$ & $4.19(3)$ & $\textcolor{gray}{0.32(3)}$ & $4.36(4)$ & $\textcolor{gray}{0.49(4)}$ & $4.64(4)$ & $\textcolor{gray}{0.60(4)}$ & $3.69(0)$ & $\textcolor{gray}{0.40(0)}$ \\[4pt]
$cs\bar{s}\bar{c}$ & $3.95(2)$ & $\textcolor{gray}{0.08(2)}$ & $4.26(3)$ & $\textcolor{gray}{0.20(3)}$ & $4.43(4)$ & $\textcolor{gray}{0.37(4)}$ & $4.71(3)$ & $\textcolor{gray}{0.50(3)}$ & $4.00(1)$ & $\textcolor{gray}{0.15(1)}$ \\[4pt]
$bn\bar{n}\bar{b}$ & $10.38(2)$ & $-0.24(2)$ & $11.27(9)$ & $\textcolor{gray}{0.59(9)}$ & $10.97(5)$ & $\textcolor{gray}{0.29(5)}$ & $11.71(8)$ & $\textcolor{gray}{0.60(8)}$ & $10.09(3)$ & $-0.71(3)$ \\[4pt]
$bs\bar{s}\bar{b}$ & $10.41(2)$ & $-0.44(2)$ & $11.30(9)$ & $\textcolor{gray}{0.39(9)}$ & $11.02(1)$ & $\textcolor{gray}{0.11(1)}$ & $11.73(7)$ & $\textcolor{gray}{0.40(7)}$ & $10.30(4)$ & $-0.64(4)$ \\[4pt]
$bc\bar{c}\bar{b}$ & $11.26(2)$ & $-1.21(2)$ & $12.06(5)$ & $-0.48(5)$ & $11.79(1)$ & $-0.75(1)$ & $12.46(6)$ & $-0.43(6)$ & $12.0(1)$ & $-0.6(1)$ \\[4pt]
\end{tabular}
\caption{\label{tab: excited masses} Same as in Table~\ref{tab: masses} but for the first radially excited states.}
\end{table*}

Starting with the $1^{++}$ channel, we find that the $cn\bar{n}\bar{c}$ ground state nicely agrees with the experimental state $\chi_{c1}(3872)$.
On the other hand, the $cs\bar{s}\bar{c}$ ground state is too light compared to the $\chi_{c1}(4140)$, while
the $cn\bar{n}\bar{c}$ and $cs\bar{s}\bar{c}$ radial excitations are in the right mass region to be identified
either with the $\chi_{c1}(4140)$ or  $\chi_{c1}(4274)$.

In the $1^{+-}$ channel, the $cn\bar{n}\bar{c}$ ground state is close to the   $Z_c(3900)$
and the $cs\bar{s}\bar{c}$ ground state  close to the $X(4020)^\pm$.
Their first radial excitations might be candidates for the $Z_c(4200)^+$ and $Z_c(4430)$,
although their masses lie substantially above the respective thresholds and should thus be treated with caution.

In the $0^{++}$ channel, we find that the $cn\bar{n}\bar{c}$ ground state agrees with the
 $\chi_{c0}(1P)$, which in the literature, however, is considered as a $c\bar{c}$ ground state.
In addition, also the $cs\bar{s}\bar{c}$ ground state appears in the same mass region.
The excited $cn\bar{n}\bar{c}$ state
is in good agreement with the $\chi_{c0}(3915)$ and the excited $cs\bar{s}\bar{c}$ state matches very nicely
with the recently observed $X(3960)$ \cite{LHCbCollaboration2022}.

In the $1^{--}$ vector channel we find that the $cn\bar{n}\bar{c}$ ground state agrees with the $\psi(4230)$, thus rendering
it the lowest-lying four-quark candidate in this channel with the dominant physical component  being $D\bar{D}_1$,
followed by $\chi_{c0}\omega$ and $J/\psi\sigma$.
The  states below the $\psi(4230)$ are not picked up by our analysis as they feature different decay channels.
As a caveat, we note that our calculated $D_1$  is substantially lighter than its experimental counterpart,
which also lowers the $D\bar{D}_1$ threshold so that our state is far above the threshold
whereas the experimental $\psi(4230)$ is a shallow bound state.
The corresponding $cs\bar{s}\bar{c}$ ground state is close to the $\psi(4360)$, although
an identification may be questionable as the dominant decays of the experimental $\psi(4360)$
do not point towards  $s\bar{s}$ components~\cite{Workman2022}.
The $cs\bar{s}\bar{c}$ excited state is however close to the $\psi(4660)$,
which because of its prominent decays to $D_s\bar{D}_{s1}(2536)$ and $\psi(2S)\pi\pi$ ($\psi(2S)f_0(980)$) is assumed to be a
hidden-charm, hidden-strange four-quark state. This leaves the $cn\bar{n}\bar{c}$ excited state, which appears to be missing from the
experimental spectrum.

The pseudoscalar $0^{-+}$ channel features the physical components $\chi_{c0}\eta$, $\eta_{c}f_0$ and
$D^\ast\bar{D}_1$. It should be kept in mind that in our present truncation the $\eta$ only features
an $n\bar{n}$ component and is thus mass-degenerate with the pion (cf. Table \ref{tab: meson-masses}).
We therefore expect this component to be possibly too dominant in our current calculation as compared
to a more complete approach. In the $\eta_c$ hadro-charmonium component we chose the $f_0(1370)$ as
companion state, since the $\sigma$ is itself a four-quark state \cite{Pelaez:2015qba,Eichmann2016a}
and too broad to act as companion. Finally, for the heavy-light meson components we did not consider
the $DD_0$ combination, because the experimentally measured $D_0$ is again much too broad to form a
molecular bound state as already argued in Refs.~\cite{Cleven2015,Brambilla2020}. Instead, we
consider the combination $D^\ast\bar{D}_1$.

As a result, we obtain the $cn\bar{n}\bar{c}$ ground state at a mass
of about $3.37$~GeV and a corresponding $cs\bar{s}\bar{c}$ ground state mass at $3.68$~GeV. We also compared our result for the $cn\bar{n}\bar{c}$ ground and excited state masses to the one using the masses for the $\eta$ and $\chi_{c0}$ given in the PDG. We found that both masses increase about $200$~MeV with respect to the masses quoted in Table~\ref{tab: masses}. However, using the PDG masses leads to a significantly higher threshold in the $\chi_{c0}\eta$ channel of about $3.96$~GeV, rendering both states deeply bound.
The masses for the corresponding excited $cn\bar{n}\bar{c}$ and $cs\bar{s}\bar{c}$ states are
 $3.61$~GeV and $4.00$~GeV, respectively.

Considering the lowest-lying heavy-light meson-meson $S$-wave thresholds for each channel we investigated, one can
identify the following threshold hierarchy. The threshold in the scalar channel is the lightest, followed by the
axialvectors, the vector and finally the pseudoscalar channel thresholds. Our states, even including other
components than only heavy-light meson-meson pairings, follow this pattern except for the pseudoscalar channel, which is fully hadro-quarkonium dominated. We come back to this point  in Section \ref{subsec: internal} below when we
discuss the internal structure of our states.

\subsection{Bottom spectrum}\label{subsec: bottom}

Moving on to the bottomonium spectrum  in the right panel of Fig.~\ref{fig: spectrum}, we observe similar
features as in the charmonium spectrum. Due to the plateau-like behaviour of the MECs for small quark masses,
the $cn\bar{n}\bar{c}$ and $cs\bar{s}\bar{c}$ states (green and blue boxes, respectively) are even closer to
each other compared to the charmonium spectrum and  overlap in most cases.

In the bottomonium spectrum there are only two experimentally well-established four-quark candidates, namely
the $Z_b(10610)$ and $Z_b(10650)$ with quantum numbers $1^{+-}$. These are potential members of an isospin
triplet and therefore indistinguishable in our isospin symmetric framework. We find a $bn\bar{n}\bar{b}$ ground
state with a slightly lower mass than the experimental $Z_b(10610)$ which seems to match reasonably well. However,
we also find a corresponding state with $bs\bar{s}\bar{b}$ flavour content close by, which has not yet been
detected in experiment.

The predicted bottomonium partners of the charmonium-like $0^{++}$, $1^{++}$ and $2^{++}$ four-quark states
are in the literature referred to as $W_{bJ}$ states. In the $1^{++}$ channel, our $bn\bar{n}\bar{b}$ ground
state coincides with the experimental state $\chi_{b1}(3P)$, which is a radial excitation of the $\chi_{b1}(1P)$.
We therefore predict a $W_{b1}$ state with a mass of about $m_{W_{b1}} = 10.52(2)$ GeV.
In the scalar channel, our $bn\bar{n}\bar{b}$ ground state is close to the $\chi_{b0}(1P)$,
as was the case in the charmonium spectrum. If the pattern that the first excited state is more in line with
the experimental four-quark candidates in this particular channel can be carried over from the charm to the
bottomonium spectrum, we predict a $W_{b0}$ with a mass of $m_{W_{b0}} = 10.38(2)$ GeV.
We note that the scalar $0^{++}$ and axialvector $1^{++}$ $W_{bJ}$ masses using heavy-quark spin symmetry (HQSS)
and effective field theories are predicted in a similar region~\cite{Baru2019}, except that they are resonances
above the respective $B\bar{B}$ and $B\bar{B}^\ast$ thresholds while our states are below their thresholds.

Considering the spectrum in the vector channel, we find the ground state with quark content $bn\bar{n}\bar{b}$
in the vicinity of three states, i.e., $\psi(10753)$, $\psi(10860)$ and $\psi(11020)$. The latter two
are considered to be radial excitations of the $\Upsilon(1S)$ often termed $\Upsilon(5S)$ and $\Upsilon(6S)$
in the literature. Therefore, despite the higher mass of our $bn\bar{n}\bar{b}$ ground state an identification
with the experimental $\psi(10753)$ seems to be an option.

Finally, for the pseudoscalar channel we find the $bn\bar{n}\bar{b}$ ground state at a mass of $9.9$~GeV being slightly heavier than our obtained $bn\bar{n}\bar{b}$ scalar state mass. Also here, substituting our pseudoscalar $n\bar{n}\, (\eta)$ mass with the $\eta$ mass from the PDG yields a $bn\bar{n}\bar{b}$ ground state which is about $200$~MeV heavier.

The $bs\bar{s}\bar{b}$ and $bc\bar{c}\bar{b}$ states in each channel do not currently have any experimental candidates.
We can, however, compare our findings to predictions from the literature, especially regarding the $bc\bar{c}\bar{b}$
states. These  have been investigated using various methods such as augmented QCD sum rules~\cite{Yang2021a},
diquark-antidiquark models~\cite{Jalili2023} and lattice-QCD inspired quark models~\cite{Yang2021}. The results
for the investigated channels with these methods lie mostly above $12$ GeV and are very close or above the respective
meson-meson thresholds. From the MECs in Figs.~\ref{fig: old}--\ref{fig: new} and the resulting
binding energies in Tables~\ref{tab: masses} and \ref{tab: excited masses} one can clearly see that our states get
more deeply bound if we increase the mass of the $q\bar{q}$ pair, i.e., when we go from $bn\bar{n}\bar{b}$
to $bs\bar{s}\bar{b}$ and $bc\bar{c}\bar{b}$. Thus, as expected, we find the $bc\bar{c}\bar{b}$ ground
states (and even their first excited states) to be deeply bound in every channel.

\subsection{Internal structure}\label{subsec: internal}

One of the most interesting questions on four-quark states concerns their internal structure.
In the preceding works using functional methods~\cite{Wallbott2019,Wallbott2020,Santowsky2022}
such information was extracted from the MECs only. The
mass spectrum of four-quark states with quark content $cq\bar{q}\bar{c}$ was calculated
by keeping the mass of the $c\bar{c}$ pair fixed and varying the mass of the light
current-quark pair $q\bar{q}$ from up/down to charm.
By changing the physical components entering in the calculation,
one can then observe how well the MEC for a single sub-cluster
(or combinations of sub-clusters) agrees with the MEC of the full state.
For example, for the $\chi_{c1}(3872)$ the $D\bar{D}^\ast$ component alone agrees
reasonably well with the full result across a wide range
of current-quark masses, while the $J/\psi\,\omega$ cluster contributes marginally
and the effect of the $SA$ diquark cluster is almost negligible (cf.~Fig.~2 in \cite{Wallbott2019}).

Here we employ a different strategy to obtain information about the internal structure
of four-quark states. To this end, we investigate the  dressing functions $f_i(S_0)$
in Eq.~(\ref{eq: dirac_bse_w_poles}) and in particular their norm contributions, similarly to
Refs.~\cite{Eichmann2022,Torcato2023,Liu:2022nku} where the orbital angular momentum composition
of baryons and the strengths of different internal diquark clusters were quantified along the same lines.
We follow
the canonical normalization procedure for BSEs \cite{Cutkosky1964,Nakanishi1965}
by contracting the amplitude in Eq.~(\ref{eq: bse_physical})
with its charge conjugate $\bar{\Gamma}^{(\mu)}$. Because the
 amplitude sums over the three dominant internal structures
(\textit{cf.} Fig.~\ref{fig: physical_bse}), the product
$\bar{\Gamma}^{(\mu)} \,G_0^{(4)}\,\Gamma^{(\mu)}$ is a sum of nine terms
as illustrated
in Fig.~\ref{fig: normcontributions}. The diagonal terms represent
the contributions coming  from the three topologies
$\mathcal{M}_1$, $\mathcal{M}_2$ and $\mathcal{D}$, and
the off-diagonal terms arise from
the mixing of these topologies.

\begin{figure}
\includegraphics[scale=0.6]{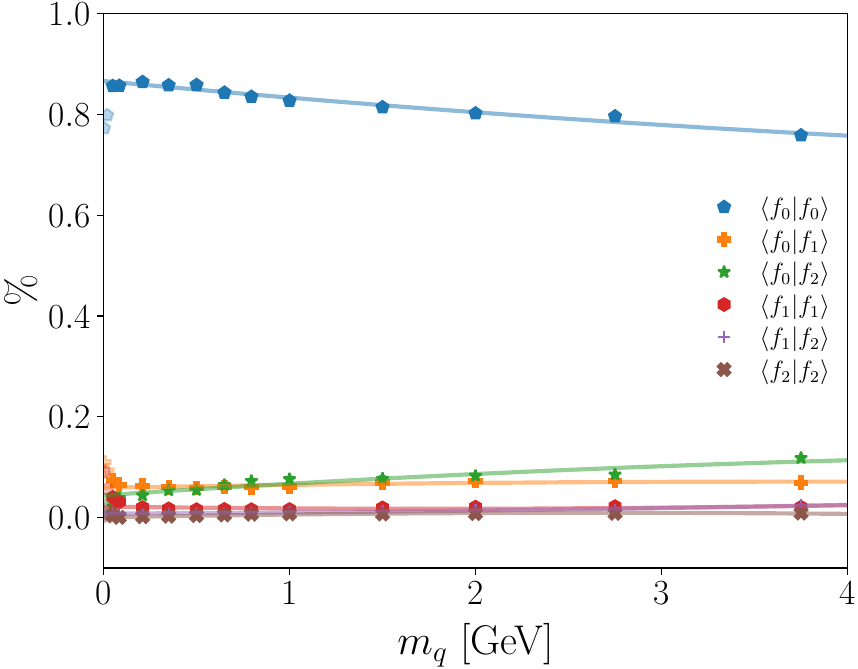}
\caption{\label{fig: norm_evc}Current-quark mass evolution of the norm contributions for the $cq\bar{q}\bar{c}$ ground states in the $0(1^{++})$ channel. The colour coding is the same as in Fig.~\ref{fig: normcontributions}.}
\end{figure}

\begin{figure}[t]
\includegraphics[scale=0.95]{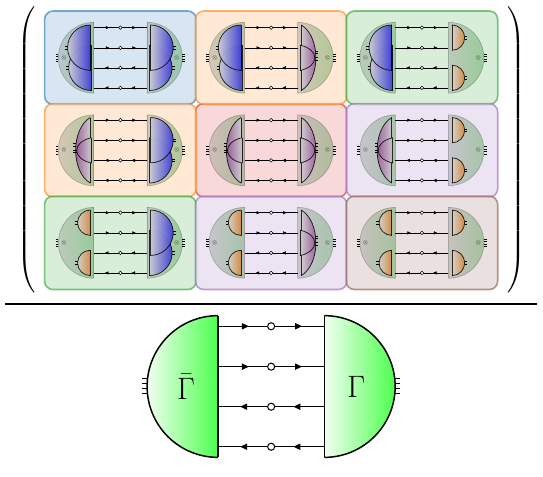}
\caption{\label{fig: normcontributions}Graphical illustration of the norm contribution matrix. Each entry in the matrix
is an overlap integral that contributes to the normalization of the four-quark state shown in the denominator.
The diagonal terms correspond to the norm contribution coming  from the $\mathcal{M}_1$, $\mathcal{M}_2$ and $\mathcal{D}$
topologies and the off-diagonal terms arise from the mixing of the topologies. Note that
the matrix is symmetric, so that the contributions with the same background colour are summed up.
}
\end{figure}

To make statements about the internal structure of the physical four-quark states 
with quark content $cn\bar{n}\bar{c}$ and $bn\bar{n}\bar{b}$, we need to calculate the 
norm contributions with the on-shell Bethe-Salpeter amplitude for that quark configuration. 
However, for some channels it is not possible to use the on-shell amplitude directly due to the two-body thresholds.
We can, however, calculate the norm contributions for a four-quark state with quark content $Qq\bar{q}\bar{Q}$ 
with the on-shell BSAs at quark masses for $q$ where the two-body thresholds do not affect the state, 
cf. Figs.~\ref{fig: old},~\ref{fig: old-1} and~\ref{fig: new}. This yields a quark mass evolution 
of the norm contributions, in analogy to the MECs, which is then extrapolated to the physical quark content. 
An example for the $0(1^{++})$ channel is shown in Fig.~\ref{fig: norm_evc}. 
One can see that the evolution follows a clear trend and does not change drastically when varying the quark masses. 
The opaque datapoints at small current-quark masses are the norm contributions not considered in the fit 
and correspond to the opaque points in Fig.~\ref{fig: old}. 
The curves for the other quantum numbers and the excited states behave in a similar way.

\begin{figure*}[t]
\raisebox{+0.45\height}{\includegraphics[width=0.35\textwidth]{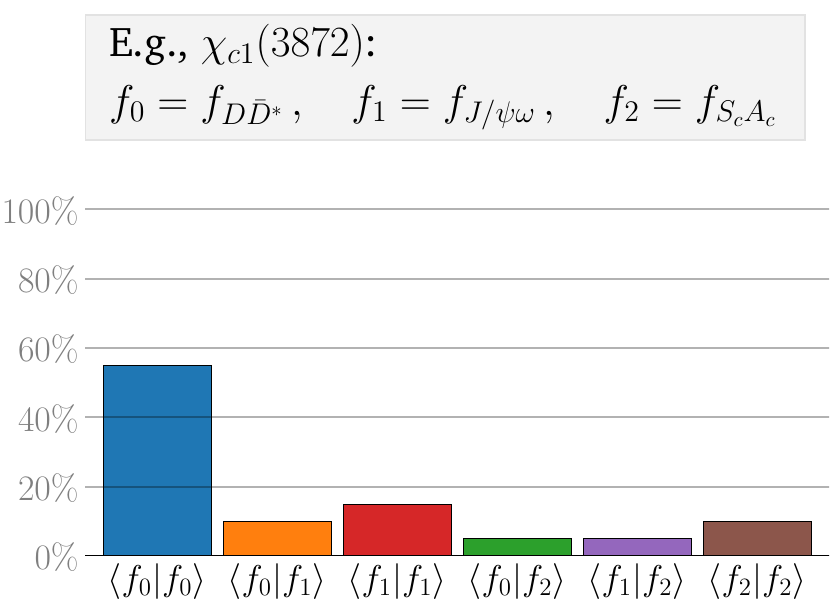}}\hfill \includegraphics[width=0.6\textwidth]{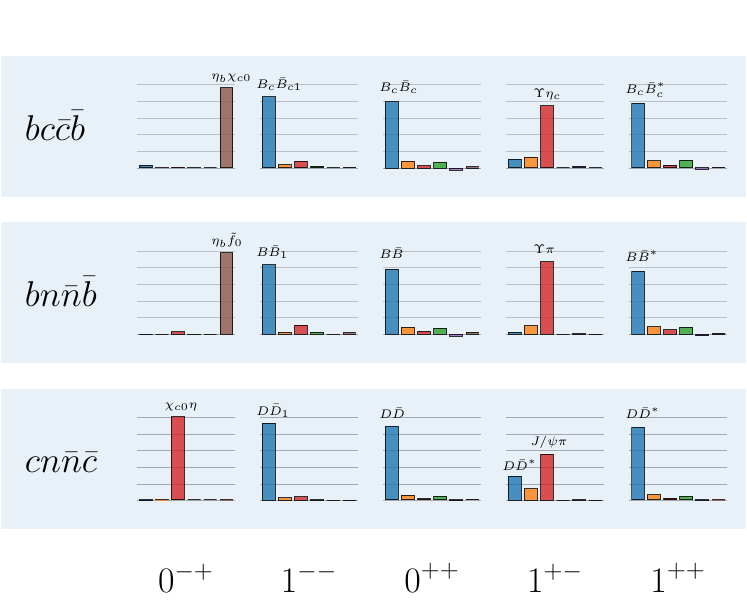}

\caption{\label{fig: norm} Norm contributions for the $cn\bar{n}\bar{c}$, $bn\bar{n}\bar{b}$ and $bc\bar{c}\bar{b}$ states.
In the scalar channel we  show the results for the excited states and in all other channels those for the ground states. The $\tilde{f}_0$ here corresponds to the $f_0(1370)$.
The bars sum up to $100\%$.}
\vspace{-1mm}
\end{figure*}

Fig.~\ref{fig: norm} shows the results for the contributions of this correlation matrix elements to each state.
The dressing functions $f_1$, $f_2$ and $f_3$ correspond
to the first, second and third physical component in Table~\ref{tab: components} for each quantum number,
e.g., $D\bar{D}$, $J/\psi\omega$ and $S_c S_c$ for the $0(0^{++})$ state.
The  plot on the left illustrates the arrangement and colour scheme.
For the $0^{-+}$, $1^{--}$, $1^{+-}$ and $1^{++}$ channels we only show the contributions for the $cn\bar{n}\bar{c}$, $bn\bar{n}\bar{b}$ and $bc\bar{c}\bar{b}$ ground states,
since they hardly change for the corresponding excited states or states with hidden strangeness ($cs\bar{s}\bar{c}$ and $bs\bar{s}\bar{b}$).
In the scalar channel we plot the results for the respective excited states as they are more in line with the exotic candidates.

Starting with the $1^{++}$ channel in the charmonium sector, we find that this state has an overwhelming $D\bar{D}^\ast$ component
(\textit{blue}) which contributes about $88\%$ to the state. The $J/\psi\omega$ component (\textit{red}) and its
mixing with $D\bar{D}^\ast$ (\textit{orange}) are almost negligible, but these are still bigger than the
$S_c A_c$ diquark component (\textit{brown}) and its mixing with  $D\bar{D}^\ast$  (\textit{green}).
This nicely reproduces  the hierarchy found in~\cite{Wallbott2019,Santowsky2022}. It also mirrors
the experimentally known decays: the dominant hadronic decay channel for the $\chi_{c1}(3872)$ is $D\bar{D}^\ast$
with $\sim 86\%$ when combining the $D^0\bar{D}^0\pi^0$ and $D^0\bar{D}^{\ast 0}$ channels,
followed by $J/\psi\omega$ with  $\sim 8\%$ \cite{Workman2022}.
Furthermore, as the $\chi_{c1}(3872)$ is very close to the $D\bar{D}^\ast$ threshold, a strong $D\bar{D}^\ast$ component in its wave function is expected.
The same behaviour is also found for the
$cs\bar{s}\bar{c}$ ground and excited state, which are dominated by the $D_s\bar{D}_s^\ast$ component.
In the bottom sector we find the (would-be $W_{b1}$) $bn\bar{n}\bar{b}$  state to be dominated by the
$B\bar{B}^\ast$ component with about $77\%$; here the other components are still all below $10\%$ but
the mixing of $B\bar{B}^\ast-\Upsilon\omega$ and $B\bar{B}^\ast-S_b A_b$ becomes
more prominent.

In the $1^{+-}$ channel we find the $J/\psi\pi$ component to be dominant ($55\%$), however with a
substantial $D\bar{D}^\ast$ admixture ($29\%$) and a $D\bar{D}^\ast-J/\psi\pi$ mixing component ($14\%$).
In the literature the
internal structure of the $Z_c(3900)$ is  debated as its mass is close to, but still
above, the $D\bar{D}^\ast$ threshold. Furthermore, the $D\bar{D}^\ast$ decay channel is preferred over
$J/\psi\pi$ by a factor $\sim 6$~\cite{Ablikim2014}.
This led to the conclusion that the
$J/\psi\pi$ component might be suppressed and the $Z_c(3900)$ could be explained as a $D\bar{D}^\ast$ molecule.
The  HAL-QCD lattice collaboration studied the internal structure of the the $Z_c(3900)$ using a $D\bar{D}^\ast-J/\psi\pi$
coupled-channel analysis~\cite{Ikeda2016,Ikeda2018}. They found a
strong $D\bar{D}^\ast-J/\psi\pi$ mixing potential and thus evidence against the state having only a
meson-meson or hadro-charmonium structure. Instead, they concluded that this strong potential leads to a formation
of the $Z_c(3900)$ as a threshold cusp. An analysis of the experimental data with effective field
theories~\cite{Chen2023,Yan2023} suggests that the meson-molecule $D\bar{D}^\ast$ component is at least
equally important than non-molecular structures.

Interestingly, the contributions in the $1^{+-}$ channel change in the bottom
sector. Here we find a very strong $\Upsilon\pi$ component ($86\%$) and a $\Upsilon\pi-B\bar{B}^\ast$
mixing of about $10\%$ for the $bn\bar{n}\bar{b}$ state, with the other correlations being negligible.
For the $bc\bar{c}\bar{b}$ state the $\Upsilon\eta_c$ component is weaker ($\sim 75\%$)
and the  $B_c\bar{B}_c^\ast$ and $\Upsilon\eta_c-B_c\bar{B}_c^\ast$ mixing contribute about $10\%$ to the state.
This is somewhat at odds with the common picture in the literature, where
the $Z_b(10610)$ is considered as a $B\bar{B}^\ast$ molecule because of the dominant decay
$Z_b(10610)\to B^+\bar{B}^{\ast 0}+B^{\ast +}\bar{B}^0$ with about $86\%$~\cite{Workman2022} and
the closeness to the $B\bar{B}^\ast$ threshold~\cite{Brambilla2020}.

In the $0^{++}$ channel we find a dominant $D\bar{D}$ component for the $cn\bar{n}\bar{c}$ state which
contributes about $89\%$ to the state.
The mixing components of $D\bar{D}-J/\psi\omega$ and $D\bar{D}-S_c S_c$ amount to about $9\%$,
while the remaining contributions are negligible. This again nicely confirms
the findings of \cite{Wallbott2020,Santowsky2022} where this state was found to be predominantly $D\bar{D}$. The corresponding
$cs\bar{s}\bar{c}$ state has a contribution of $D_s\bar{D}_s$ of about $89\%$ which fits with the observed decay
channel of the $X(3960)$ \cite{LHCbCollaboration2022}. In the bottom region, we  observe again a $B\bar{B}$
dominance of $80\%$ with the mixings $B\bar{B}-\Upsilon\omega$ and $B\bar{B}-S_b S_b$ becoming more prominent
with a total contribution of $16\%$.

Concerning the $cn\bar{n}\bar{c}$ state in the $1^{--}$ vector channel, we find that it is almost
exclusively dominated by the
 $D\bar{D}_1$ component ($93\%$) with a small contribution coming from the $\chi_{c0}\omega$
($4\%$) and $D\bar{D}_1-\chi_{c0}\omega$ mixing ($3\%$). This is in line with Refs.~\cite{Detten2023,vonDetten:2024eie}
concluding that a description of the $\psi(4230)$ as a $D_1\bar{D}$ molecular state agrees with the
experimental data. As the $D\bar{D}_1$ threshold is also the lowest $S$-wave meson-meson threshold in the
system~\cite{Hanhart2017,Guo2018,Brambilla2020}, the closeness of the state to that threshold also points
to a strong meson-molecule component in the wave function. The internal structure of the $\psi(4660)$, which
we identified with our $cs\bar{s}\bar{c}$ excited state, is a little more elaborate. Here, motivated by the
observed decays, there are claims to describe this state as a hadro-charmonium $\psi(2S)f_0(980)$ state~\cite{Guo2008},
a $D_s^{(\ast)}\bar{D}_{s1}(2536)$ meson-molecule~\cite{Wang2020}, or a $P$-wave tetraquark ($\mathit{dq-\overline{dq}}$)
state \cite{Zhang2020}. In our analysis we find the $D_s\bar{D}_{s1}$ component to be dominant with $92\%$.
The picture is quite similar in the bottom region: All states are dominated by the
respective $B\bar{B}_1$ component with about $83\%$, followed by a contribution of about $10\%$ coming from
the pure $\chi_{b0}\omega$ cluster.

Finally, we turn to the $0^{-+}$ pseudoscalar channel. For the $cn\bar{n}\bar{c}$ state
we find that with a dominant $\chi_{c0}\eta$ component of $99\%$ our obtained $cq\bar{q}\bar{c}$ states are
exclusively hadro-charmonium with no mixing components. As discussed, this picture might change when we
include the strange flavour components in the $\eta$ together with the octet-singlet mixing and therefore needs
to be explored again in a more complete approach in the future.
In the bottom sector the dominant substructure is still hadro-quarkonium, but the state is now almost
exclusively dominated by the $\eta_b f_0(1370)$ hadro-bottomonium component with about $95\%$, augmented by an
almost negligible $\chi_{b0}\eta$ contribution of $5\%$. Presumably this result will not change in a more complete
approach given that a heavier $\eta$ will decrease the importance of this component rather than increase it.
When going to the $bc\bar{c}\bar{b}$ state, the dominant component is still $\eta_b\chi_{c0}$ with $96\%$, with the $B^\ast B_1$ and the $\chi_{b0}\eta_c$ components contributing about $2\%$ each.

\section{Summary and conclusions}\label{conclusions}

In this work we determined the spectrum and internal composition of heavy-light four-quark states with hidden flavour
in the charm and bottom energy region. Using  rainbow-ladder two-body interactions between quarks and
(anti-)quarks, we solved the four-body Faddeev-Yakubovsky equation in a fully covariant framework and obtained spectra
for states with quantum numbers $J^{PC}=0^{++}, 0^{-+}, 1^{--}, 1^{+-}$ and $1^{++}$. Our wave functions routinely
incorporate contributions from internal heavy-light meson-meson, hadro-quarkonium and diquark-antiquark contributions.
Whereas we find the latter ones to be subleading in most cases, it turns out that states with different quantum numbers
correspond to different internal contributions, which in some cases also depend on the (hidden) flavour of the states.
For all states with $CP=+1$ (i.e. $J^{PC}=0^{++}, 1^{--}$ and $1^{++}$) we find highly dominant heavy-light meson
contributions, which almost exclusively determine the masses of the ground and excited states in the charm and bottom
energy region. For axialvector states with $J^{PC}=1^{+-}$, however, we find a much more complicated picture. Our
hidden-charm state corresponding to the experimental $Z_c(3900)$ comprises both heavy-light meson but also
hadro-charmonium components in qualitative agreement with results from other approaches
\cite{Ikeda2016,Ikeda2018,Chen2023,Yan2023}. The corresponding state in the bottom energy region is even dominated
by the hadro-bottomonium configuration. The same is observed for our pseudoscalar states regardless of flavour.
Thus the most important message from our study is: the internal composition of XYZ states is by no means uniform
but varies between different quantum numbers and flavours. It is certainly interesting to extend our findings to the
open flavour case. Corresponding work is in progress and will be reported elsewhere.

\acknowledgments{We thank Marc Wagner and the Frankfurt group for extended and very fruitful discussions on the
subject. This work was supported by the BMBF under project number 05P2021, the DFG under grant number FI 970/11-2,
the graduate school HGS-HIRe and the GSI Helmholtzzentrum f\"ur Schwerionenforschung. This work contributes to the
aims of the U.S. Department of Energy ExoHad Topical Collaboration, contract DE-SC0023598. We acknowledge computational
resources provided by the HPC Core Facility and the HRZ of the Justus-Liebig-Universit\"at Gie\ss en.}



\appendix

\section{Construction of  amplitudes}\label{app: new amplitudes}

In this section we describe the construction of the Bethe-Salpeter amplitudes for the $0(1^{--})$ vector
and $0(0^{-+})$ pseudoscalar four-quark states based on their dominant physical clusters.
We refer to the supplemental material of Ref.~\cite{Wallbott2020} for the analogous construction
 for the quantum numbers $J^{PC} = 0^{++}$, $1^{+-}$ and $1^{++}$.

To begin with, we consider the quantum numbers of the meson-meson and diquark-antidiquark components that
can appear in a four-quark state with the desired quantum numbers. These components can then be
assigned to one of the interaction topologies $\mathcal{M}_1$ and $\mathcal{M}_2$ (meson-meson) or $\mathcal{D}$ (diquark-antidiquark).
We then compare with the PDG \cite{Workman2022} for the experimentally dominant or realized decays of the four-quark
state under consideration. If there is no experimental evidence regarding the dominant decays, we take the
lowest-lying thresholds of the system as the dominant contributions spanning the physical basis.
For the vector states this procedure is straightforward as there are well-established exotic candidates with vector quantum numbers such as the
$\psi(4230)$. At present, however, there is no evidence of a pseudoscalar four-quark candidate, so
we have to determine its physical clusters based solely on the lowest-lying thresholds of the possible internal
components.

As a first approximation, we neglect the diquark components in the basis for the vector and
pseudoscalar four-quark states. The reasons for this are twofold. First, the possible diquark clusters feature
either $S$-wave pairings of scalar/axialvector diquarks with pseudoscalar or vector diquarks
(which are not only heavier than their scalar and axialvector  counterparts,
but also unreliable in a rainbow-ladder truncation), or $P$-wave scalar and axialvector diquark cluster
pairings (which include higher orbital angular momentum and are therefore suppressed). The notion of $S$ and $P$ wave here refers to the combination of Dirac tensors 
needed to construct a basis element of a four-quark state with vector or pseudoscalar quantum numbers. The $P$-wave combination would induce angular momentum between the two diquarks, which is however not possible with the approximations explained in Sec.~\ref{subsec: physical poles}.
Secondly, the diquark clusters  always form the highest threshold in the system and   were previously found to have a
negligible influence on the mass of  hidden-flavour four-quark states~\cite{Eichmann2016a,Wallbott2019,Wallbott2020,Santowsky2022a}.
The resulting physical components chosen for the vector and pseudoscalar four-quark states are then only
of meson-meson type and given in Table~\ref{tab: components}.

To construct the basis describing the pseudoscalar and vector states, we write down the leading Dirac-colour tensors
for $I(0^{-+})$ and $I(1^{--})$ in analogy to the supplemental material of~\cite{Wallbott2020}. For the pseudoscalar states this yields
\begin{align}\label{aeq: basis_pseudoscalar}
\varphi_1^{\pm} &= \big[\gamma_{\alpha\gamma}^5\delta_{\beta\delta} \pm \delta_{\alpha\gamma}\gamma_{\beta\delta}^5\big] C_{11}\nonumber\, ,\\[4pt]
\varphi_2^{\pm} &= \big[\gamma_{\alpha\delta}^5\delta_{\beta\gamma} \pm \delta_{\alpha\delta}\gamma_{\beta\gamma}^5\big] C'_{11}\nonumber \, ,\\[4pt]
\varphi_3^{\pm} &= \big[\big(\gamma_\perp^\mu\big)_{\alpha\gamma}\big(\gamma^5\gamma_\perp^\mu\big)_{\beta\delta} \pm \big(\gamma^5\gamma_\perp^\mu\big)_{\alpha\gamma}\big(\gamma_\perp^\mu\big)_{\beta\delta}\big] C_{11}\, ,\\[4pt]
\varphi_4^{\pm} &= \big[\big(\gamma_\perp^\mu\big)_{\alpha\delta}\big(\gamma^5\gamma_\perp^\mu\big)_{\beta\gamma} \pm \big(\gamma^5\gamma_\perp^\mu\big)_{\alpha\delta}\big(\gamma_\perp^\mu\big)_{\beta\gamma}\big] C'_{11}\nonumber\, .
\end{align}
where $\gamma_\perp^\mu = (\delta^{\mu\nu}-\hat{P}^\mu\hat{P}^\nu)\gamma^\nu$ is the transverse projection of the
$\gamma$-matrices with respect to the normalized total momentum $\hat{P}^\mu$ and $C_{11} = \delta_{AC}\delta_{BD}/3$, $C'_{11} = \delta_{AD}\delta_{BC}/3$ are
the two colour-singlet tensors, with colour indices $A,B,C,D=1,2,3$. For the vector states we use
\begin{align}\label{aeq: basis_vector}
\psi_1^{\pm} &= \big[\gamma_{\alpha\gamma}^5\big(\gamma^5\gamma_\perp^\mu\big)_{\beta\delta} \pm \big(\gamma^5\gamma_\perp^\mu\big)_{\alpha\gamma}\gamma_{\beta\delta}^5 \big] C_{11} \nonumber\, ,\\[4pt]
\psi_2^{\pm} &= \big[\gamma_{\alpha\delta}^5\big(\gamma^5\gamma_\perp^\mu\big)_{\beta\gamma} \pm \big(\gamma^5\gamma_\perp^\mu\big)_{\alpha\delta}\gamma_{\beta\gamma}^5 \big] C'_{11} \nonumber\, ,\\[4pt]
\psi_3^{\pm} &= \big[\delta_{\alpha\gamma}\big(\gamma_\perp^\mu\big)_{\beta\delta} \pm \big(\gamma_\perp^\mu\big)_{\alpha\gamma}\delta_{\beta\delta}\big]C_{11}\, ,\\[4pt]
\psi_4^{\pm} &= \big[\delta_{\alpha\delta}\big(\gamma_\perp^\mu\big)_{\beta\gamma} \pm \big(\gamma_\perp^\mu\big)_{\alpha\delta}\delta_{\beta\gamma}\big] C'_{11} \nonumber \, ,
\end{align}
Since we are dealing with heavy-light hidden-flavour four-quark states, we  retain only those basis elements that fulfil charge
conjugation symmetry ($C$ parity) in the $(14)(23)$ topology:
\begin{equation}\label{aeq: basis_phys}
\{\varphi_1^+,\, \varphi_2^+,\, \varphi_2^-,\, \varphi_3^-\} \quad \text{and} \quad
\{\psi_1^-,\, \psi_3^+,\, \psi_4^+,\, \psi_4^-\}\,.
\end{equation}
Concerning the pseudoscalar basis,
we are aware of the fact that $\gamma^5\gamma_\perp^\mu$
is the leading tensor  for the $1^{++}$ two-body state, whereas the leading $1^{+-}$ meson tensor needed
to form the  quantum numbers of the four-quark state has additional higher angular momentum structures~\cite{Williams2010}.
However, for the  physical components in Table~\ref{tab: components} we can equally take the leading tensor
 for the $1^{++}$ two-body state as they are degenerate for heavy-light systems.

\begin{figure}
\includegraphics[scale=0.5]{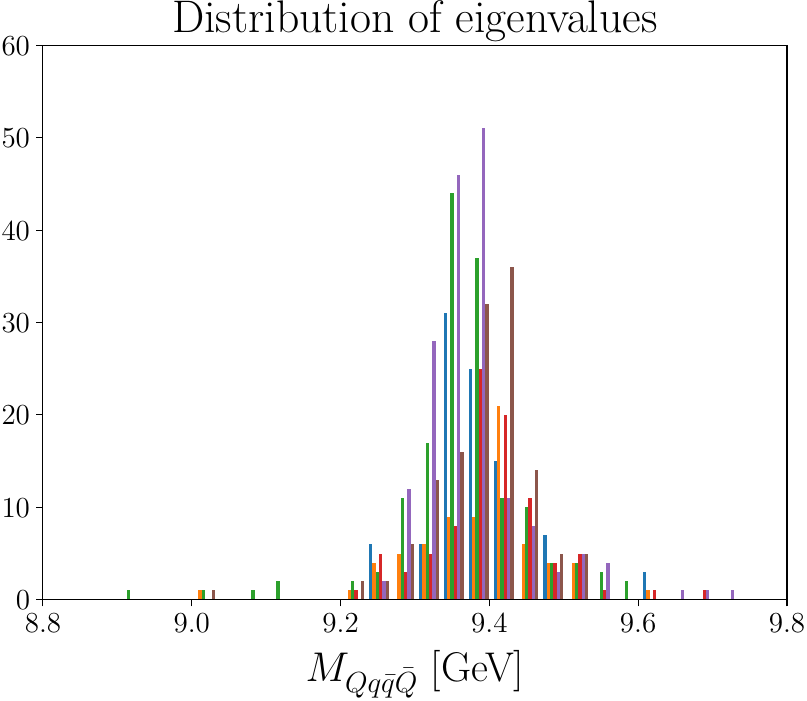}
\caption{\label{fig: Distribution}Example distribution of extrapolated eigenvalues for the $I(J^{PC})=0(1^{++})$ state
with quark configuration $cq\bar{q}\bar{c}$ and $m_q=2750$ MeV. The
extrapolation results are centred around a value of $M_{cq\bar{q}\bar{c}}\approx 9.4$ GeV with a spread that resembles
a Gaussian distribution.}
\end{figure}

\section{Error analysis}\label{subsec: error}

In model calculations it is always hard to determine a systematic error. In previous works \cite{Wallbott2019,Wallbott2020}
the main parameter $\eta^{\mathrm{MT}}=1.8$ of our model interaction, given by Equation (3.96) in \cite{Eichmann2016}                                                                                        as already noted above, has been varied by $\pm 0.2$ to gauge the effect on the masses of the four-quark states.
Since these turned out to be remarkably stable under such variations we do not repeat this exercise here.

However, we need to take into account potential errors due to the extrapolation of our eigenvalue curves into the
complex momentum plane. As discussed in Sec.~\ref{subsec: physical basis}, two-body thresholds can affect the                                                                                                calculation such that the mass of the four-quark state cannot be obtained directly. Therefore, we have to resort to                                                                                          extrapolation of the eigenvalues above the threshold. For this we use a variation of the Schlessinger-Point-Method (SPM)
that works as follows:
Having obtained a set of eigenvalues $\mathcal{A}_r = \{\lambda(P_i^2)\}_{i=1}^r$,
we extrapolate the behaviour of all $r$ eigenvalues as a function of $P^2$ to the value where the condition
$\lambda(P_i^2=-M_i^2)=1$ is fulfilled. This gives us a base estimate of the mass, i.e., $M_{\mathrm{base}}$,
and the first element in a set of extrapolated masses denoted by $\mathcal{B}$. Next we choose a random subset
of eigenvalues $\mathcal{A}_m\subset \mathcal{A}_r$, with $m\in\{r-1,r-2,r-3,r-4,r-5,r-6\}$ and extrapolate
the eigenvalues chosen in $\mathcal{A}_m$ to fulfil the condition again. This procedure is repeated about $300$
times for each $m$. The corresponding extrapolation results are added to the set $\mathcal{B}$ if they lie in
a $5\%$ region around the $M_{\mathrm{base}}$ value. To obtain the masses shown in Figs.~\ref{fig: old}--\ref{fig: new} we average the values in the set $\mathcal{B}$ and the error is given by the standard deviation
of the values in $\mathcal{B}$.
A nice example of the distribution of eigenvalues in $\mathcal{B}$ is shown
in the histogram in Figure~\ref{fig: Distribution}.
This method has also been applied to states where the condition $\lambda_i=1$ can be read off from the eigenvalue
curve directly. The corresponding histogram plot of eigenvalues then shows an extreme dense clustering around
the directly determined value, which is the expected behaviour.

\section{First radially excited states}\label{subsec: excited_states}
\begin{figure*}[!b]
	\includegraphics[scale=0.54]{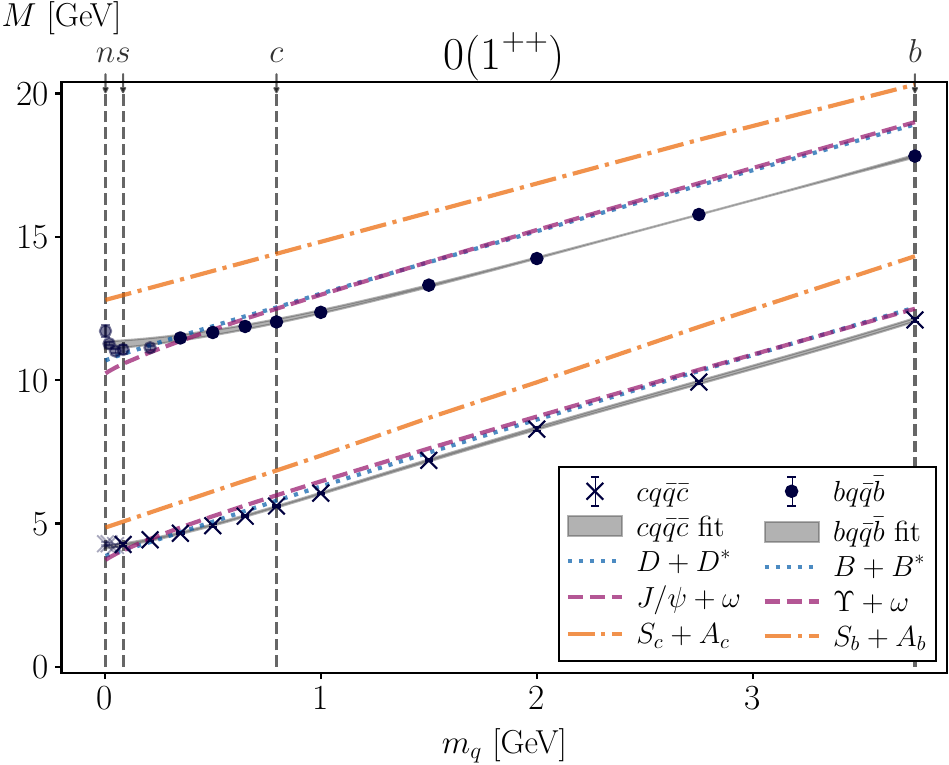}							\hfill\includegraphics[scale=0.54]{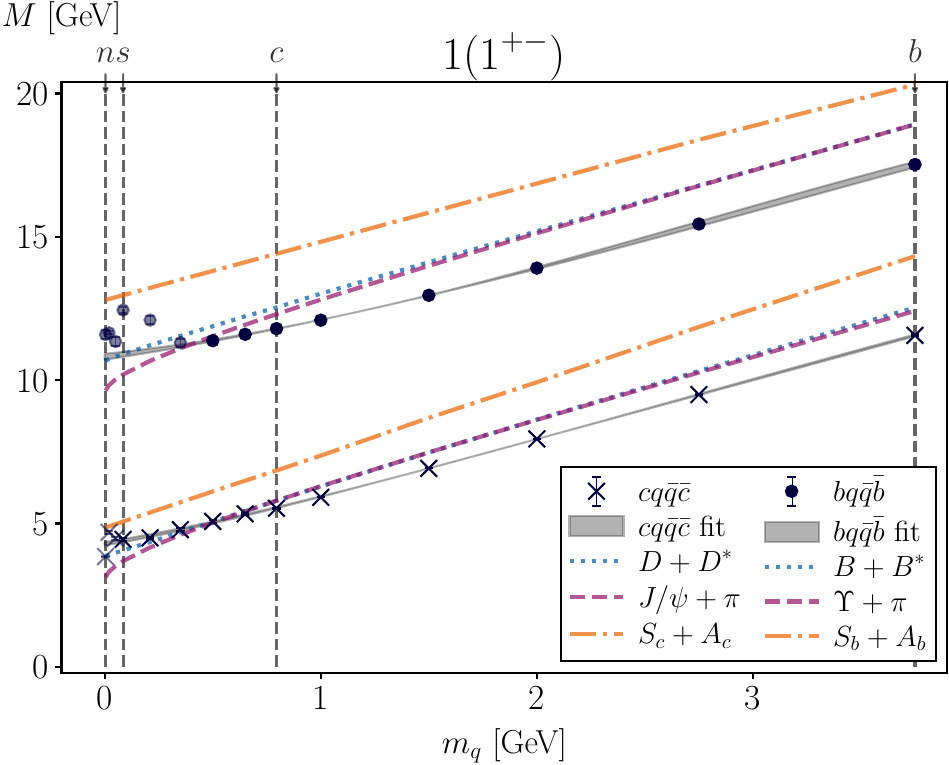}\\
	\includegraphics[scale=0.54]{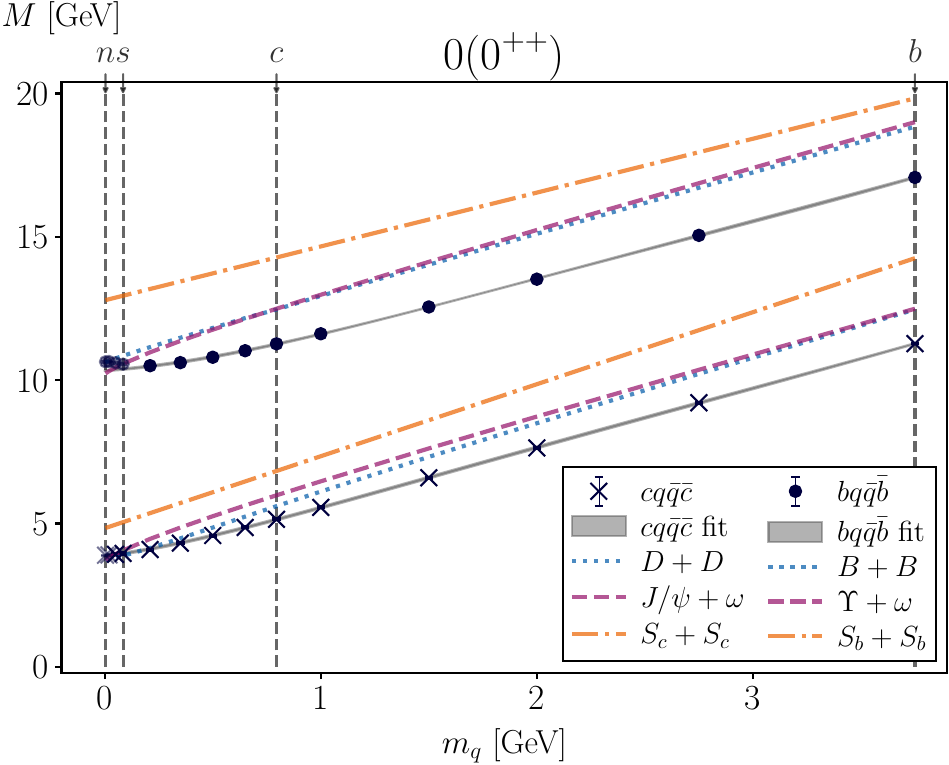}\hfill \includegraphics[scale=0.54]{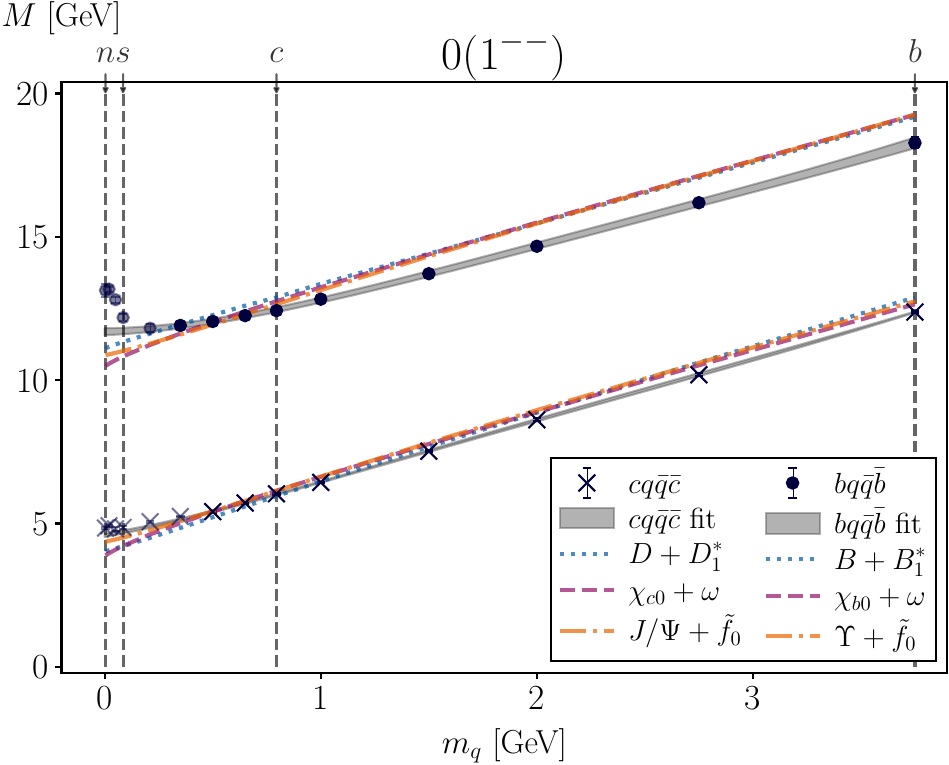}\\
\includegraphics[scale=0.54]{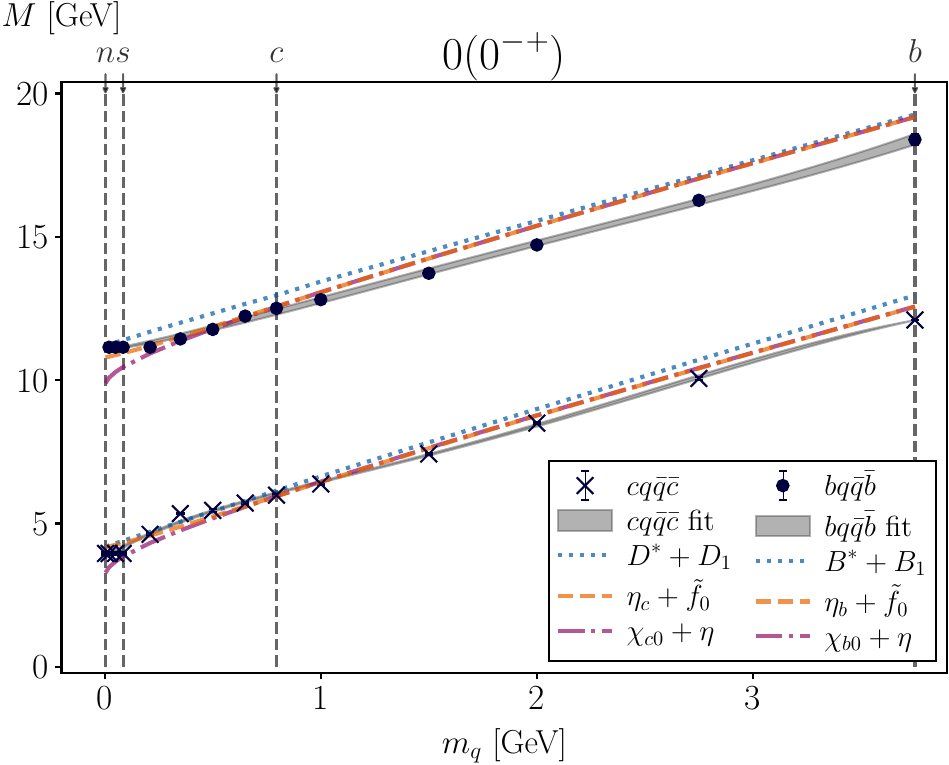}
	\caption{\label{fig: old_ex} Current-mass evolution of the $cq\bar{q}\bar{c}$ (crosses) and $bq\bar{q}\bar{b}$ first radial excited states (dots)
		for the investigated quantum numbers in this work; see Fig.~\ref{fig: old} for details.}
\end{figure*}
In Figs.~\ref{fig: old_ex} we show the MECs for the first radial excitations, which yield the first excited state masses in Table~\ref{tab: excited masses}. Comparing these MECs to the ones from Figs.~\ref{fig: old},~\ref{fig: old-1} and~\ref{fig: new}, we see that the curves are affected by threshold effects much earlier than the curves for the ground states. Therefore, the number of datapoints which can be directly calculated, and thus fitted by the fit given in Eq.~(\ref{eq: fit}), is reduced compared to the ground state MECs.

\clearpage

\bibliography{hidden_flavour}

\end{document}